# Coating Breakdown Prediction for Ships and Inspection Planning

Huy Truong-Ba[1,*], Michael E. Cholette[1], Geoffrey Will[2], Marc Hartmann[3]

1. *Queensland University of Technology, Brisbane Queensland, Australia*
2. *University of Sunshine Coast, Australia*
3. *Thales Group, Australia*

**Corresponding author:* h.truongba@qut.edu.au

## Abstract

Marine corrosion significantly reduces a ship's availability, increases costs of operation and could impact safety. Protective coatings mitigate these risks, but their effectiveness deteriorates over time. Early detection of coating breakdown is crucial to prevent costly repairs and safety concerns. While corrosion itself is well-understood, coating degradation remains under-investigated due to insufficient long-term data. This work addresses this knowledge gap by enhancing coating defect prediction and optimizing inspection planning for ships. The Power Law Non-Homogeneous Poisson Process (PL-NHPP) is utilized for modeling coating defect arrivals. Unlike prior studies, we employ a hierarchical Bayesian approach for parameter fitting, effectively addressing limitations associated with scarce real-world data. Furthermore, we optimize inspection planning by incorporating out-of-service costs and potential costs increases due to delayed repairs. The efficacy of these methods is evaluated through a comprehensive case study involving a recently commissioned fleet with limited historical data. This research contributes to the advancement of condition-based maintenance (CBM) strategies for ships by enabling more accurate prediction of coating breakdowns and optimizing inspection schedules early in the life of the fleet. This approach ultimately improves operational efficiency and reduces life-cycle costs.

## 1 Introduction

Corrosion poses a significant threat to ships, progressively reducing their operational lifespan and ultimately limiting their functionality. This degradation process can lead to structural failures and safety concerns. To mitigate these risks, protective coatings are applied to ship surfaces, creating a barrier against the corrosive marine environment (Eccles et al., 2010).

However, the effectiveness of these coatings diminishes over time due to various factors such as ultraviolet radiation, mechanical wear, and the chemical environment (Melchers and Jiang, 2006). Early detection of coating breakdown, the point where the coating loses its protective function, is crucial to prevent the onset of corrosion and the associated costly repairs (Abbas and Shafiee, 2020).

While extensive research has focused on understanding and mitigating corrosion processes themselves (Luque et al., 2014; Yazdi et al., 2021), the degradation of the protective coating layer remains a relatively under-investigated area (Davies et al., 2021; Melchers, 2008). One of the reason is because of the lack of long-term coating data under controlled conditions (Melchers, 2008; Melchers and Jiang, 2006). Additionally, traditional maintenance relies on inspections to identify coating failure, which is a reactive approach and necessitates the need for models to predict coating breakdown. These models are essential for establishing effective inspection regimes for different ships that are specific to individual compartments.

A scarcity of unbiased, long-term field data hinders comprehensive research on the combined effects of protective coating degradation and subsequent corrosion defect arrival (i.e., coating breakdowns) (Melchers, 2008; Melchers and Jiang, 2006). Despite this challenge, efforts have been made to address this knowledge gap. (Melchers and Jiang, 2006) attempted an initial estimation of coating life in water ballast tanks through expert surveys encompassing vessel users, coating contractors, and suppliers. (Nicolai et al., 2007) modeled the random degradation process of coatings on steel structures using Brownian and Gamma processes. (Momber et al., 2022) furthered this research by introducing a coating degradation function specifically for onshore wind turbines. For describing coating defect arrivals, the Poisson Process has emerged as a powerful tool (Castro et al., 2011; Davies et al., 2021; Kuniewski et al., 2009; Nicolai et al., 2007). Within this framework, the non-homogeneous Poisson process (NHPP) is the most frequently employed model, with the Power Law – NHPP (PL-NHPP) being the most common variant (Davies et al., 2021; Kuniewski et al., 2009; Nicolai et al., 2007).

Accurate prediction of coating defect arrivals hinge not only on selecting an appropriate model, but also on effectively fitting the model's parameters, especially when real-world data is available. Maximum Likelihood Estimation (MLE) is a commonly used method for parameter estimation in ship studies (Davies et al., 2021; Nicolai et al., 2007). However, MLE encounters limitations when dealing with sparse data sets. In our previous work (Davies et al., 2021), a data grouping approach was proposed to address this limitation by combining data from similar ship compartments, thereby enriching the data for parameter fitting. An alternative approach to overcome data sparsity is the Bayesian framework (Rios Insua et al., 2012).

Notably, this approach has not yet been explored in the context of coating breakdown prediction.

Ship corrosion maintenance is transitioning towards condition-based maintenance (CBM) strategies. CBM prioritizes inspections to identify corroded components, enabling targeted repairs (Dong and Frangopol, 2015). However, the effectiveness of CBM hinges on careful inspection planning to prevent severe corrosion damage. Ideally, inspections should be condition-based, driven by the real-time monitored condition of the ship (Ai et al., 2020; Truong-Ba et al., 2021). This approach, however, necessitates extensive measurements of coating thickness loss across all ship components (Luque et al., 2014), rendering it impractical for large-scale implementation. Consequently, time-based inspection planning remains the prevailing approach in both industry practice and the current literature (Davies et al., 2021).

This work builds upon our previous study (Davies et al., 2021) to enhance coating defect prediction and optimize inspection planning for ships. We achieve this through the following contributions:

- We employ the Power Law Poisson Non-Homogeneous Poisson Process (PL-NHPP) for modeling coating defect arrivals. However, unlike our previous work, we utilize a hierarchical Bayesian approach for parameter fitting, effectively addressing limitations arising from scarce data.
- We optimize the inspection planning process by incorporating out-of-service costs associated with harboring the ship for inspections and potential cost increases due to delayed repairs of coating defects.
- To evaluate the efficacy of the developed methods and models, we conduct a comprehensive analysis using a real-world case study involving a recently commissioned fleet with limited data.

This paper describes the parameter estimation procedures for the PL-NHPP model using Bayesian approaches, with a specific focus on addressing the challenges associated with limited data through a Hierarchical Bayesian statistics. We present the development of inspection planning optimization models for determining intervals and scheduling. The results obtained by fitting the defect arrival model to real-world data from operational vessels are discussed and a sensitivity analyses is conducted.

## 2 Ship Coating Breakdown Arrival Model

### 2.1 Non-homogeneous Poisson Process (NHPP) model for Ship Coating Breakdown Arrival Times

The deterioration of protective coatings with ship age leads to an increased frequency of coating defects. The Power Law Non-Homogeneous Poisson Process (PL-NHPP) is a well-established method for modeling this phenomenon. This incorporates an arrival intensity that reflects the time-dependent nature of defect occurrences (Babishin and Taghipour, 2016; Nicolai et al., 2007; Yang et al., 2017). As shown in the following equation:

$$\lambda(t) = abt^{b-1} \tag{1}$$

Where $a$ and $b$ are the characteristic life and shape parameters, respectively. Notably, a higher $b$ value indicates a stronger correlation between increasing ship age and defect arrival rate. Conversely, $\mathrm{b} = 1$ signifies an age-independent arrival rate. Let $N_{t_1,t_2}$ represents the random number of defects observed within a specific time interval $[t_1, t_2]$ in some area of interest (e.g. a compartment). Under the PL-NHPP model, the number of defect arrivals has the probability distribution:

$$f\left(N_{t_1,t_2}; a, b\right) = \Pr\left[N_{t_1,t_2}\right] = \frac{\Lambda(t_1, t_2)^{N_{t_1,t_2}}}{N_{t_1,t_2}!} e^{-\Lambda(t_1,t_2)} \tag{2}$$

where $\Lambda(t_1, t_2)$ is the cumulative intensity function:

$$\Lambda(t_1, t_2) = \int_{t_1}^{t_2} \lambda(t) = a\left(t_2^b - t_1^b\right) \tag{3}$$

This quantity is also the expectation of defects between two time points:

$$\mathbb{E}\left[N_{t_1,t_2}\right] = \Lambda(t_1, t_2) \tag{4}$$

### 2.2 Fitting the NHPP model

Spatial heterogeneity in exposure conditions and protective measures can lead to significant variations in coating degradation across different ship compartments. To address this spatial variability, we categorize coating defect data based on ship and compartment dimensions. Coating breakdowns are typically identified and reported during routine inspections. As a consequence of the potentially varying breakdown rates among compartments, optimal inspection intervals are established for each. However, a key limitation of the available data is the absence of precise defect arrival times. Instead, the data captures the number of defects observed within a compartment during the interval between inspections. These defect counts are subsequently employed to estimate the parameters $(a, b)$ of the aforementioned PL-NHPP model.

Parameter estimation for the PL-NHPP model typically involves maximizing the log-likelihood function of the model with respect to the available data. This data reflects the number of coating breakdowns observed within inspection intervals for each compartment. A specific compartment is represented as $j$ on a ship $s$, and $k$ denotes the time interval between two consecutive inspections $\left[t_{s,j,k-1}, t_{s,j,k}\right)$ $\forall k = 1,2 \ldots$ The number of coating defects observed within this compartment during this interval $k$ is denoted as $N_{s,j,k}$. Hence, the data structure for coating breakdown arrival times can be formally defined as:

$$\mathcal{D}_{s,j} = \left\{\left(t_{s,j,1}, N_{s,j,1}\right), \left(t_{s,j,2}, N_{s,j,2}\right), \ldots\right\}$$

And the data set for all ships, denoted as $\mathcal{D}$ is defined as:

$$\begin{aligned} \mathcal{D} &= \left\{\mathcal{D}_{s,j}\ ; s = 1..N_s\ , j = 1..N_c\right\} \\ &= \left\{\left(t_{s,j,k}, n_{s,j,k}\right); \ = 1..N_s\ , j = 1..N_c, k = 1..K_j\right\} \end{aligned}$$

According to this data structure and characteristics of PL-NHPP model, the log-likelihood function has a form of Poisson distribution:

$$\begin{aligned} \ell\left(a, b \middle| \mathcal{D}_{s,j}\right) &= \sum_{k=1}^{K} -\Lambda\left(t_{s,j,k-1}, t_{s,j,k}; a, b\right) + N_{s,j,k} \ln \Lambda\left(t_{s,j,k-1}, t_{s,j,k}; a, b\right) - \ln N_{s,j,k}! \\ &= \sum_{k=1}^{K} -a\left(t_{s,j,k}^{b} - t_{s,j,k-1}^{b}\right) + N_{s,j,k} \ln a + N_{s,j,k} \ln\left(t_{s,j,k}^{b} - t_{s,j,k-1}^{b}\right) \\ &\quad - \ln N_{s,j,k}! \end{aligned} \tag{5}$$

where $\Lambda\left(t_{s,j,k-1}, t_{s,j,k}; a, b\right)$ is the cumulative intensity parameter given by $a$ and $b$. Due to fact that $\ln N_{s,j,k}!$ is constant for given $\mathcal{D}_{s,j}$, the parameter estimation $\left(\hat{a}_{s,j}, \hat{b}_{s,j}\right)$ of compartment $j$ of ship $s$ is determined by:

$$\begin{aligned} \left(\hat{a}_{s,j}, \hat{b}_{s,j}\right) &= \operatorname*{argmax}_{a,b} \ell\left(a, b \middle| \mathcal{D}_{s,j}\right) \\ &\equiv \operatorname*{argmax}_{a,b} \sum_{k=1}^{K} -\Lambda\left(t_{s,j,k-1}, t_{s,j,k}; a, b\right) + N_{s,j,k} \ln \Lambda\left(t_{s,j,k-1}, t_{s,j,k}; a, b\right) \\ &= \operatorname*{argmax}_{a,b} \sum_{k=1}^{K} -a\left(t_{s,j,k}^{b} - t_{s,j,k-1}^{b}\right) + N_{s,j,k} \ln a + N_{s,j,k} \ln\left(t_{s,j,k}^{b} - t_{s,j,k-1}^{b}\right) \end{aligned} \tag{6}$$

While the Maximum Likelihood Estimation (MLE) method offers a straightforward approach to parameter estimation, it has limitations when applied to the PL-NHPP model with limited data. Notably, MLE requires at least three data points for parameter estimation, and its characterization of parameter uncertainty suffers in sparse data scenarios. This can lead to unreliable confidence intervals. Additionally, MLE solely relies on the data itself, neglecting the potential value of incorporating prior engineering knowledge about the parameters. To address these limitations and leverage such knowledge, the Bayesian approach emerges as a valuable alternative.

### 2.2.1 Fitting PL-NHPP model with Bayesian approach

In this method, the parameter set $\theta = (a, b)$ of PL-NHPP model is assumed to be a random variable following a *prior* probability density function denoted as $p(\theta)$. This prior encapsulates all available knowledge about the parameters before seeing the data (e.g. the parameter must be positive). This distribution is combined with the data according to well-known Bayes' Rule:

$$p(\theta|\mathcal{D}) = \frac{\ell(\mathcal{D}|\theta)\, p(\theta)}{\int \ell(\mathcal{D}|\theta)\, p(\theta) d\theta} \tag{7}$$

The term $p(\theta|\mathcal{D})$ is called as *posterior* of parameters according to evidence (observation or data) $y$ and $\ell(\mathcal{D}|\theta)$ is the log-likelihood function. The denominator term $\int \ell(\mathcal{D}|\theta)\, p(\theta) d\theta$ is the expectation of evidence (or observation) given the entire domain of the parameter set and it is a constant for a given observation. Therefore, the posterior is proportional to the numerator term and denoted as:

$$p(\theta|\mathcal{D}) \propto \ell(\mathcal{D}|\theta)\, p(\theta)$$

The algebraic calculation of term $\int \ell(\mathcal{D}|\theta)\, p(\theta) d\theta$ is impossible for most practical cases, and therefore, the numerical integration can be considered, but this approach has several computational disadvantages. Another powerful way to obtain the posterior with any prior distribution is using Markov Chain Monte Carlo (MCMC) methods (Gelman et al., 2013), which is implemented in standard software such as PyMC in Python (Fonnesbeck et al., 2017), Stan or R (Gelman et al., 2013).

Unlike the MLE approach, which yields a single point estimate for the parameters, the Bayesian approach provides a posterior distribution (or a sample drawn from it) of the entire parameter set. This distribution incorporates both prior knowledge about the parameters and the observed data. Consequently, the Bayesian framework offers a more nuanced understanding of the impact of limited data and the value of incorporating engineering expertise. Notably, even with minimal data, the Bayesian approach can still produce a sample of parameter sets. In the extreme case of no data, this sample would be drawn entirely from the prior distribution, reflecting the available knowledge before data collection.

For parameter estimation within the PL-NHPP model applied to individual compartments across the entire fleet, the log-likelihood function $\ell(\mathcal{D}|\theta)$, as presented in Eq. (5), can be expressed as $\ell(\mathcal{D}|\theta) \triangleq \ell(a,b|\mathcal{D}_{s,j})\ \forall s,j$. Here, both the characteristic parameter $a$ and $b$ are inherently positive within the PL-NHPP model. To facilitate the application of informative prior distributions for these parameters, we employ their natural logarithm values, denoted as $(\ln a, \ln b)$, respectively. These transformed parameters can range from negative to positive infinity. In contrast, if we assume that coating breakdowns across all compartments and ships exhibit similar behavior and can be modeled by a single PL-NHPP model, all data points can be pooled for parameter estimation. Consequently, the resulting log-likelihood function takes the following form:

$$\ell(a,b|\mathcal{D}) = \sum_{s=1}^{N_s}\sum_{j=1}^{N_c}\sum_{k=1}^{K_j}\Big[-\Lambda\big(t_{s,j,k-1}, t_{s,j,k}; a, b\big) + N_{s,j,k}\ln\Lambda\big(t_{s,j,k-1}, t_{s,j,k}; a, b\big) - \ln N_{s,j,k}!\Big] \tag{8}$$

The selection of priors for the model parameters $(\ln a, \ln b)$ can be driven by the knowledge and experience of ship operators. When operators or ship experts possess insights or prior beliefs regarding the parameter ranges, informative priors can be employed. These informative priors, such as Normal or Log-Normal distributions, should be accompanied by appropriate standard deviations to reflect the level of certainty in the prior knowledge. Conversely, in situations where there is limited or no specific knowledge about the distribution of coating breakdown arrivals, or their occurrence is highly uncertain, non-informative priors with a wide range, such as a flat prior, are preferred.

### 2.2.2 Hierarchical Bayesian approach for partial data pooling

The previous section explored parameter fitting approaches for the coating breakdown prediction model in two contrasting scenarios: (1) a single, universal parameter set for all compartments across all ships, and (2) unique parameter sets for each individual compartment on every ship. Both approaches present challenges. Individual fitting offers an unbiased solution, particularly when substantial coating failure data is available. However, real-world data, especially regarding defects, is often limited. Consequently, a more nuanced approach is necessary. This could involve partial grouping of compartments that share similar degradation mechanisms, such as grouping compartments of the same type across different ships. For Bayesian inference, a powerful technique known as Hierarchical Bayesian modeling can be employed to partially pool data from similar compartments. In this hierarchical framework, prior parameters for each compartment are treated as samples drawn from a common hyperprior distribution. The degree of data pooling is controlled by specifying the parameters of these hyperpriors, also known as hyperparameters.

The general framework of a Hierarchical Bayesian model can be described as follows. We denote $y_i$ as an observation (i.e. data) generated from a stochastic model with parameter $\theta_i$, $p(y_i|\theta_i)$. All parameters $\theta_i$ are assumed to follow a common prior distribution $p(\theta_i|\phi)$ with hyperparameter $\phi$. This hyperparameter $\phi$ itself has a hyperprior distribution $p(\phi)$. Hence, the Bayesian hierarchical structure introduces three levels (one more than a typical Bayesian model) to capture the relationships between parameters and hyperparameters:

- Stage 1: $y_i \sim p(y_i|\theta_i)$
- Stage 2: $\theta_i \sim p(\theta_i|\phi)$
- Stage 3: $\phi \sim p(\phi)$

According to Bayesian relation, we can estimate the posterior as proportion:

$$p(\theta_i|y_i) \propto p(y_i|\theta_i)p(\theta_i) = p(y_i|\theta_i)p(\theta_i|\phi)p(\phi)$$

Markov Chain Monte Carlo (MCMC) methods, which utilizes sampling processes from both the hyperprior $p(\phi)$ and prior distribution $p(\theta_i|\phi)$, can be employed to estimate the posterior distributions of the model's parameters. These posterior distributions provide valuable information regarding the uncertainty associated with each parameter.

In this study, the hierarchical fitting approach is set with parameters as follows:

- ***Priors***: the parameters for PL-NHPP models are $(\ln a, \ln b)$, which priors are still Normal distributions:

$$\ln a \sim \mathcal{N}\left(\mu_{\ln a}, \sigma_{\ln a}^2\right) \text{ and } \ln b \sim \mathcal{N}\left(\mu_{\ln b}, \sigma_{\ln b}^2\right)$$

- ***Hyperpriors***: The hyperparameters (i.e. parameters of priors) follows common hyperprior distributions:

$$\mu_{\ln a} \sim \mathcal{N}\left(m_{\ln a}, s_{\ln a}^2\right) \text{ and } \sigma_{\ln a} \sim \mathcal{U}(l_{\ln a}, u_{\ln a})$$
$$\mu_{\ln b} \sim \mathcal{N}\left(m_{\ln b}, s_{\ln b}^2\right) \text{ and } \sigma_{\ln b} \sim \mathcal{U}(l_{\ln b}, u_{\ln b})$$

# 3 Inspection Planning

To mitigate the detrimental effects of corrosion, ships undergo regular and systematic inspections of their various sections and compartments. Identified corrosion defects are promptly documented for repair and prioritized for rectification. Stringent fleet management policies mandate the comprehensive repair of all detected defects within a compartment in the allowed timeframes according to their priorities. Consequently, the accumulation of numerous corrosion defects significantly hinders vessel operations, highlighting the importance of robust preventive measures.

Furthermore, the appearance of coating defects signifies the initiation of a progressive corrosion process that worsens over time. This ongoing corrosion creates a cascade of problems, making repairs increasingly complex, time-consuming, and expensive. Therefore, the time defects remain undetected between inspections is crucial and need to be considered in inspection planning to prevent the escalation of coating-related issues. This critical factor significantly influences the strategic development of inspection protocols.

## 3.1 Number of defects and existing time of defects in inspection intervals

Let $D_{s,m}(t_1, t_2)$ represent the number of defects observed within the time interval $(t_1, t_2]$. Based on the PL-NHPP model, the expected number of defects during this interval, denoted by $\bar{D}_{s,m}(t_1, t_2)$, is calculated as shown in Eq. (4) above. Thus, the expected total number of defects accumulated over the entire lifespan of the ship (i.e. the considered time horizon) can be determined as:

$$\bar{D}_{s,m}(t_{now}, t_{end}) = \sum_{i_{s,m}=1}^{I_{s,m}} \bar{D}_{s,m}\left(t_{i_{s,m}-1}, t_{i_{s,m}}\right) = \sum_{i_{s,m}=1}^{I_{s,m}} a_{s,m}\left(t_{i_{s,m}}^{b_{s,m}} - t_{i_{s,m}-1}^{b_{s,m}}\right) \tag{9}$$

with $\quad t_{i_{s,m}} > t_{now}$ and $t_{I_{s,m}} \leq t_{end} < t_{I_{s,m}+1}$

Where $i_{s,m}$ and $t_{i_{s,m}}$ represent the index and associating time of inspection of compartment $m$ of ship $s$ from current time to the end of considered time horizon; $I_{s,m}$ denotes the total number of inspections until the end of the time horizon. Equation (9) is valid under the assumption that defects identified during a specific inspection are subsequently repaired and, therefore, will not be recorded in subsequent inspections.

Within the context of a ship compartment undergoing scheduled inspections, defects can manifest at stochastic time points. We denote the random arrival time of a defect occurring between two successive inspections $(t_1, t_2]$ as $T_k, t_1 \leq T_k \leq t_2$. The age of the defect at the time of inspection, denoted by $A_k = t_2 - T_k$, represents the duration of its presence within the compartment and is taken as a proxy for the (unobserved) defect severity. Given the arrival rate of defect is $\lambda(t)$ with cumulative intensity function $\Lambda(t_1, t)$ as PL-NHPP model above, the arrival of $k^{th}$ defect follows Gamma distribution with CDF and PDF as follows (Ross, 2014; Truong Ba et al., 2017):

CDF: $$F_{k,t_1}(t) = \frac{\gamma(k, \Lambda(t_1, t))}{(k-1)!} = \frac{\gamma\left(k, a\left(t^b - t_1^b\right)\right)}{(k-1)!} \tag{10}$$

PDF: $$f_{k,t_1}(t) = \frac{[\Lambda(t_1, t)]^{k-1}}{(k-1)!}\lambda(t)e^{-\Lambda(t_1,t)} = \frac{\left[a\left(t^b - t_1^b\right)\right]^{k-1}}{(k-1)!}(abt^{b-1})e^{-a\left(t^b - t_1^b\right)} \tag{11}$$

where $\gamma(k, s)$ is the lower incomplete gamma function.

The expected age of $k^{th}$ defect, $\bar{A}_k(t_1, t_2)$ is calculated as follows:

$$\begin{aligned}\bar{A}_k(t_1, t_2) &= \int_{t_1}^{t_2} (t_2 - t) f_{k,t_1}(t)dt = t_2 F_{k,t_1}(t_2) - \int_{t_1}^{t_2} t f_{k,t_1}(t)dt \\ &= \frac{t_2 \gamma\left(k, a\left(t_2^b - t_1^b\right)\right)}{(k-1)!} \\ &- \frac{e^{at_1^b}}{a^{\frac{1}{b}}(k-1)!}\sum_{i=0}^{k-1}\binom{k-1}{i}\left(-at_1^b\right)^i\left[\gamma\left(k - i + \frac{1}{b}, at_2^b\right)\right. \\ &\left. - \gamma\left(k - i + \frac{1}{b}, at_1^b\right)\right]\end{aligned} \tag{12}$$

The calculation for Eq. (12) is detailed in the Appendix.

## 3.2 Cost model

The development of an optimization model for inspection planning necessitates a robust cost model to quantify the economic implications of various inspection strategies. This study leverages the cost model established by (Davies et al., 2021) while incorporating targeted modifications to capture the influence of defect age on overall cost. The proposed cost model encompasses two key cost components:

- *Compartment-Specific Cost*: This component reflects the costs associated with inspecting individual compartments within a ship, including any potential maintenance activities necessitated by identified defects.
- *Ship Out-of-Service Cost*: This component captures the economic impact of the ship being unavailable for operation due to scheduled inspections and any subsequent maintenance work.

In level of compartment, the cost for each compartment $m$ of each ship $s$ at a specific inspection $i_{s,m}$ is comprised of the inspection cost for the compartment and the repair cost that relates to the age of defects:

$$C_{m,s,i_{s,m}} = c_{m,s}^{ins} + \sum_{\ell=1}^{\infty} f_c\left[\bar{A}_\ell\left(t_{i_{s,m}-1}, t_{i_{s,m}}\right)\right] \tag{13}$$

Where $c_{m,s}^{ins}$ is the fixed inspection cost for compartment $m$ of ship $s$; and $f_c(a)$ is the function of repair cost that relies on defect age $a$. In the term of expected age of defects, i.e. $\sum_{\ell=1}^{\infty} \bar{A}_\ell\left(t_{i_{s,m}-1}, t_{i_{s,m}}\right)$, when $\ell$ increased the $\bar{A}_\ell\left(t_{i_{s,m}-1}, t_{i_{s,m}}\right)$ become smaller. Hence this summation is calculated until the value of $k$ that makes $\bar{A}_\ell\left(t_{i_{s,m}-1}, t_{i_{s,m}}\right)$ is sufficiently small.

Determining the precise relationship between repair cost and defect age requires in-depth analysis of corrosion mechanisms and their progression. Such analyses are beyond the scope of this study. Therefore, for our purposes, we adopt a simplified power-law relationship between repair cost $c_{m,s,\ell}^{rep}$ and defect age $a_\ell$ as shown in Eq. (14):

$$c_{m,s,\ell}^{rep} = f_c(a_\ell) = \alpha_{m,s} a_\ell^{\beta_{m,s}} \tag{14}$$

In the repair cost function, if parameter $\beta_{m,s} = 1$, This signifies a linear relationship between defect age and repair cost. In this scenario, the cost increases proportionally with the defect's age. In case $\beta_{m,s} > 1$, this represents an exponential relationship, implying a more rapid increase in repair cost as the defect ages. This reflects the potential for greater deterioration and complexity of repair for older defects.

Considering a time horizon from 0 to time $t_K$ of the ship lifetime. Let $t_k$ $(k = 0,1,2,\dots,K-1)$ be the time candidates for ship inspection and maintenance. In practice, durations between two successive time $t_k, t_{k+1}$ are assumed to be constant as $\Delta t$. Let $x_{m,s,k} \in \{0,1\}$ be the decision variable of selecting compartment $m$ of ship $s$ for inspection at time candidate $t_k$, in which 1 means selected and 0 is otherwise. In addition, we also define the set of selected compartments for inspection in specific time $t_k$ as $\mathcal{M}_{s,k} = \left\{m \middle| x_{m,s,k} = 1\right\}$. Finally, the total cost of inspection and maintenance of ship can be determined as:

$$\begin{aligned} C_{s,k}\left(\mathcal{M}_{s,k}\right) &= c_s \cdot \max\left\{x_{m,s,k} \middle| \forall m\right\} + \sum_{m=1}^{N_m} x_{m,s,k} \cdot C_{m,s,k} \\ &= c_s \cdot \max\left(x_{m,s,k}\right) + \sum_{m=1}^{N_m} x_{m,s,k} \cdot \left[c_{m,s}^{ins} + \sum_{\ell=1}^{\infty} f_c\left[\bar{A}_\ell\left(t_{i_{s,m}-1}, t_{i_{s,m}}\right)\right]\right] \end{aligned} \tag{15}$$

where $c_s$ represents the costs of out-of-mission and/or call-out services of ship. $t_{i_{s,m}}$ is the last inspection time of compartment $m$ of ship $s$, which is defined as:

$$t_{i_{s,m}} = \max\left\{t_\ell \cdot x_{m,s,\ell} \middle| \ell = 0,1,2,\dots,k-1\right\} \tag{16}$$

### 3.3 Inspection Planning Optimization

The optimized inspection planning is one that can minimize the total inspection and repair cost for ship:

$$TC^*(\mathcal{M}_{s,1}, \mathcal{M}_{s,2}, \dots) = \min_{\mathcal{M}_{s,k}, k=1,2,\dots,K} \sum_{k=1}^{K} C_{s,k}(\mathcal{M}_{s,k})$$

Subject to:

$$\mathcal{M}_{s,k} = \{m | x_{m,s,k} = 1\}$$
$$x_{m,s,k} \in \{0,1\} \quad m = 1..N_m, k = 0..K \tag{17}$$
$$x_{m,s,K} = 1 \quad \forall m, s$$

Equation (17) above represents a general inspection planning model where individual compartments can be scheduled for inspection at any designated maintenance time point within the planning horizon. This flexibility allows for targeted inspections based on compartment risk profiles or historical maintenance data. It is important to note that the constraint $x_{m,s,K} = 1 \; \forall m, s$ ensures that all compartments undergo inspection by the end of the considered planning horizon. This constraint prevents the emergence of a trivial optimal solution where no inspections are conducted throughout the horizon, resulting in a cost of zero. Such a scenario wouldn't reflect realistic inspection practices and could lead to potential safety risks.

This section explores a simpler alternative to the optimization model presented earlier. This approach assumes a fixed inspection interval for each compartment within a ship. Common fixed intervals might be 6, 12, or 24 months, depending on industry standards, regulatory requirements, or the specific risks associated with different compartments. The case study presented in this work exemplifies the application of a fixed interval inspection strategy. In this context, we introduce a new decision variable, $y_{m,s}$. This variable represents the number of inspection interval units (i.e. $\Delta t$) between inspections for compartment $m$ within ship $s$. For instance, if $y_{m,s}$ is assigned a value of 2 for a fixed interval of 3 months, it signifies that compartment $m$ on ship $s$ will be inspected every 6 months. Then the variable $x_{m,s,k} \in \{0,1\}$ representing an inspection decision at time $t_k$ and set of inspected compartments $\mathcal{M}_{s,k}$ are determined as:

$$x_{m,s,k} = \mathbb{1}\left(k \,\%\, y_{m,s} = 0\right)$$
$$\mathcal{M}_{s,k} = \{m | x_{m,s,k} = 1\} = \{k \,\%\, y_{m,s} = 0\}$$

Where $\mathbb{I}(x)$ is the indicator function that gives 1 if $x$ is true and 0 for otherwise.

Finally, the optimization model in Eq. (17) is transformed to:

$$TC^{*}(y_{1,s}, y_{2,s}, \dots) = \min_{y_{m,s}, m=1,2,\dots,M} \sum_{k=1}^{K} C_{s,k}(\mathcal{M}_{s,k})$$

Subject to:

$$\mathcal{M}_{s,k} = \{m | x_{m,s,k} = 1\} \tag{18}$$

$$x_{m,s,k} = \mathbb{I}(k \,\%\, y_{m,s} = 0)$$

$$y_{m,s} \in \{1,2,\dots,K\}$$

$$x_{m,s,K} = 1 \quad \forall m, s$$

The optimization models presented in this work, particularly the general model in Eq. (17), are classified as combinatorial problems due to the vast number of compartments and potential inspection times. Solving these models using traditional Mixed-Integer Linear Programming (MILP) techniques can be computationally expensive, especially for realistic scenarios involving a large number of compartments and diverse inspection possibilities. In addition, the key parameters employed in the defect arrival model may not be readily available as a fixed set. Instead, they might be derived from samples or estimations, introducing additional complexities when formulating and solving a definitive MILP model. In light of these challenges, we propose a heuristic approach in well-known Genetic Algorithm (GA), specifically designed to address the unique problem of inspection scheduling for ships with a multitude of compartments.

The formidable scale of optimization models remains a significant challenge for Genetic Algorithm (GA) methodologies within the general optimization framework. Consequently, a heuristic approach has been devised to address this challenge. This heuristic strategy involves the classification of ship compartments into distinct groups based on their respective rates of defect occurrence. Subsequently, all compartments within each group adhere to a unified inspection schedule. This strategic approximation effectively diminishes the optimization task from managing schedules for individual compartments to handling schedules for groups, rendering the optimization model more tractable and solvable.

# 4 Case Study

## 4.1 Data available to this study

This study investigates a fleet of three marine vessels with varying launch dates. The launch date serves as the baseline (age zero) for each ship. Each vessel comprises up to 580 compartments susceptible to coating degradation. Model fitting for inspection scheduling relies on data collected from ship inspection reports. As illustrated in Figure 1, a limited number of compartments (i.e. fewer than 30) have experienced more than two defects. This scarcity of individual compartment data restricts the effectiveness of fitting processes that depend on such detailed information.

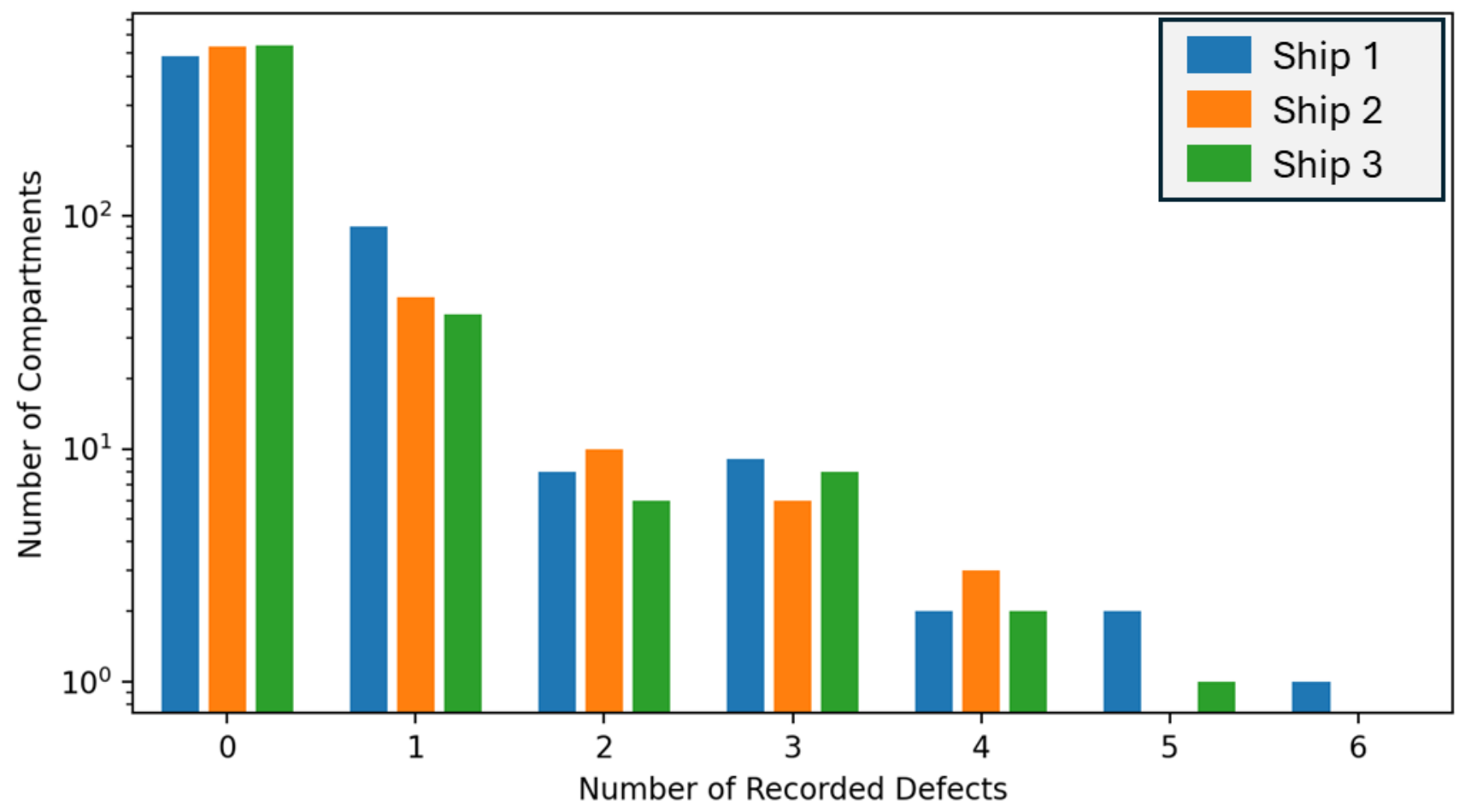

Figure 1: Histogram of number of defect reports

Due to varying rates of coating breakdown expected across compartments, each compartment is assigned a dedicated inspection interval. These intervals can be 12, 24, 30, or 60 months, resulting in a dynamic inspection schedule over time, as depicted in Figure 2. By considering the established inspection regime for individual compartments and the launch dates of the ships, it is possible to estimate the inspection schedule for each compartment relative to its launch date within the overall fleet timeline.

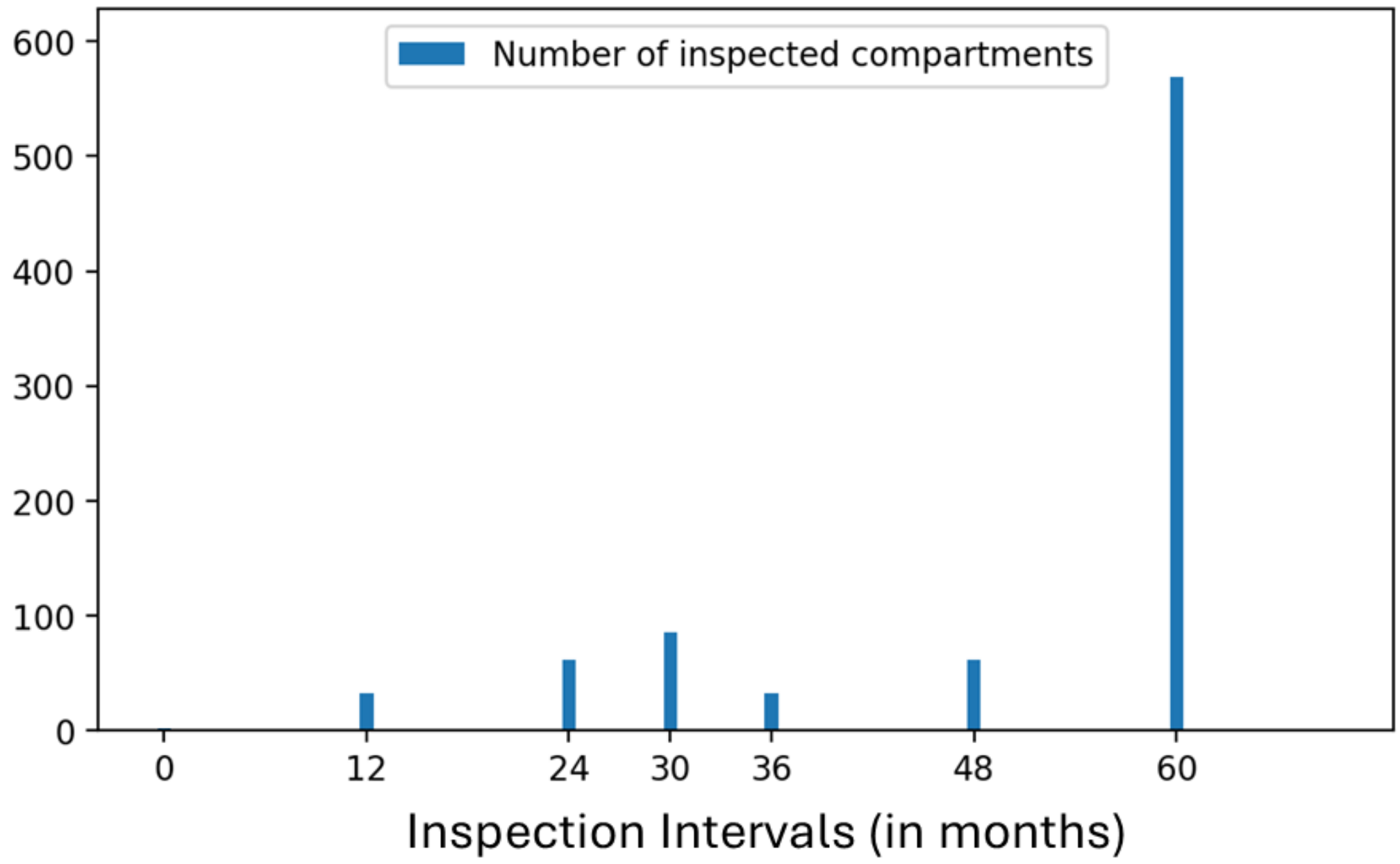

Figure 2: Number of compartments is planned for inspection at different time intervals

## 4.2 Independent fitting vs. hierarchical fitting

For this individual Bayesian fitting approach, the selected priors of $\ln a$ and $\ln b$ are Normal distributions:

$$\ln a \sim \mathcal{N}(-7,5^2) \text{ and } \ln b \sim \mathcal{N}(0,3^2)$$

For the hierarchical approach, the parameters are set as follows:

- ***Priors***: the parameters for PL-NHPP models are $(\ln a, \ln b)$, which priors are still Normal distributions:

$$\ln a \sim \mathcal{N}\left(\mu_{\ln a}, \sigma_{\ln a}^2\right) \text{ and } \ln b \sim \mathcal{N}\left(\mu_{\ln b}, \sigma_{\ln b}^2\right)$$

- ***Hyperpriors***: The hyperparameters (i.e. parameters of priors) follows common hyperprior distributions:

$$\mu_{\ln a} \sim \mathcal{N}(-7,4^2) \text{ and } \sigma_{\ln a} \sim \mathcal{U}(0,5)$$
$$\mu_{\ln b} \sim \mathcal{N}(-2,2^2) \text{ and } \sigma_{\ln b} \sim \mathcal{U}(0,3)$$

The selection of these parameters is from the statistics in hand data and preliminary fitting in which all data is used, and flat priors are set. The statistics of posteriors from preliminary fitting are then used for setting these parameters.

Figure 3 presents a comparison between individual (Bayesian) and hierarchical fitting approaches. This figure depicts the cumulative number of defects projected over the next five years for specific compartments, estimated using these distinct fitting models. Notably, the hierarchical approach demonstrates a narrower range compared to the individual approach.

This observation arises from the hierarchical model's capacity to leverage all available data in a hierarchical manner. Consequently, the hierarchical approach offers an advantage in terms of increased confidence in the fitted parameters, especially for compartments with limited data, provided that the data available exhibits sufficiently similar patterns.

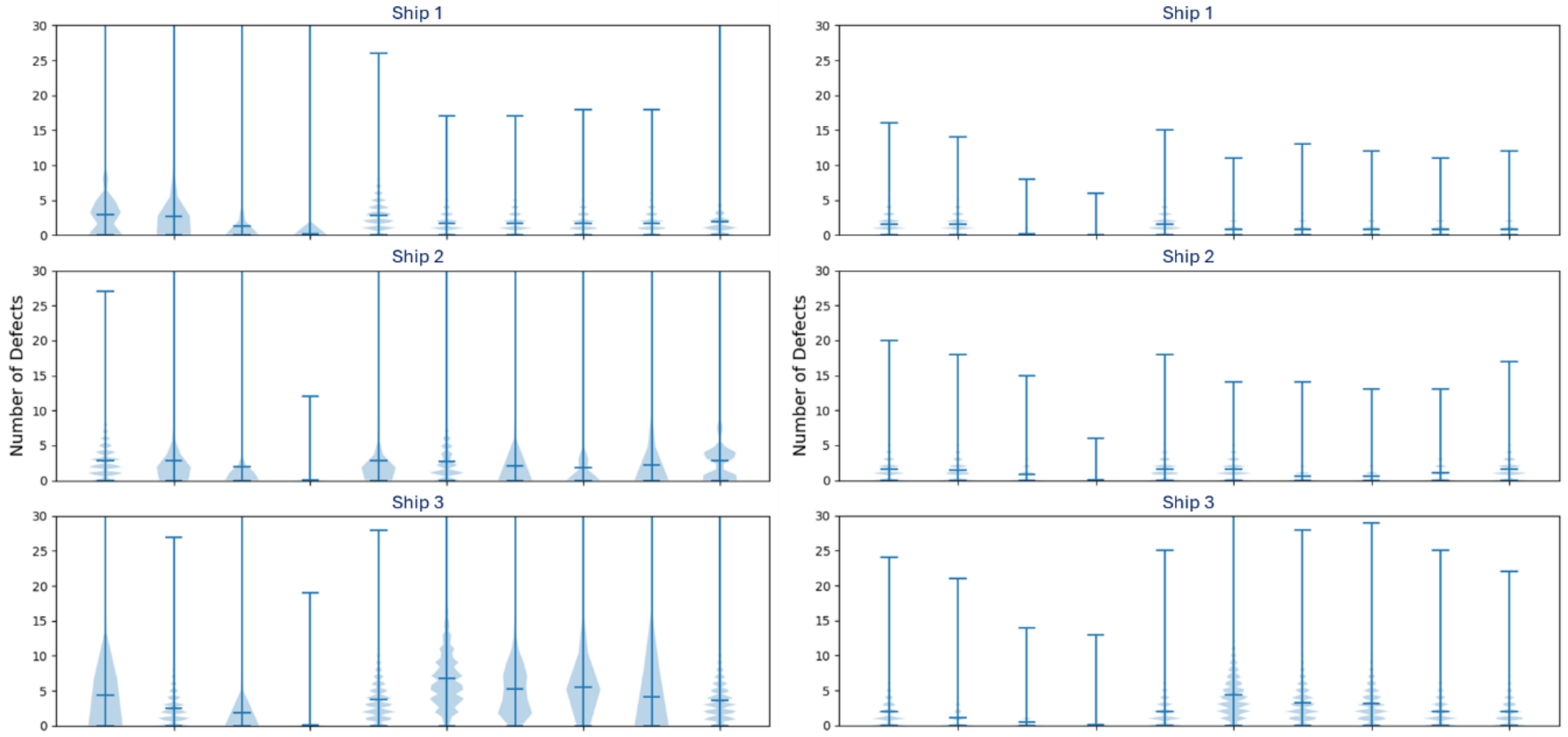

Figure 3: Number of Defects in next 5 years: Individual fitting (left) vs. Hierarchical fitting (right)

Figure 4 exemplifies the application of hierarchical Bayesian fitting for three ships, employing two distinct fitting approaches. In both scenarios – sufficient and insufficient data – the hierarchical approach achieves satisfactory fitting while exhibiting narrower confidence intervals compared to individual fitting. Notably, for cases with limited data, the rate of increase in the confidence interval for individual fitting is significantly faster than that observed in the hierarchical approach. These findings suggest that hierarchical Bayesian fitting offers a compelling set of attributes for estimating parameters within coating breakdown arrival models, particularly when data scarcity is a concern.

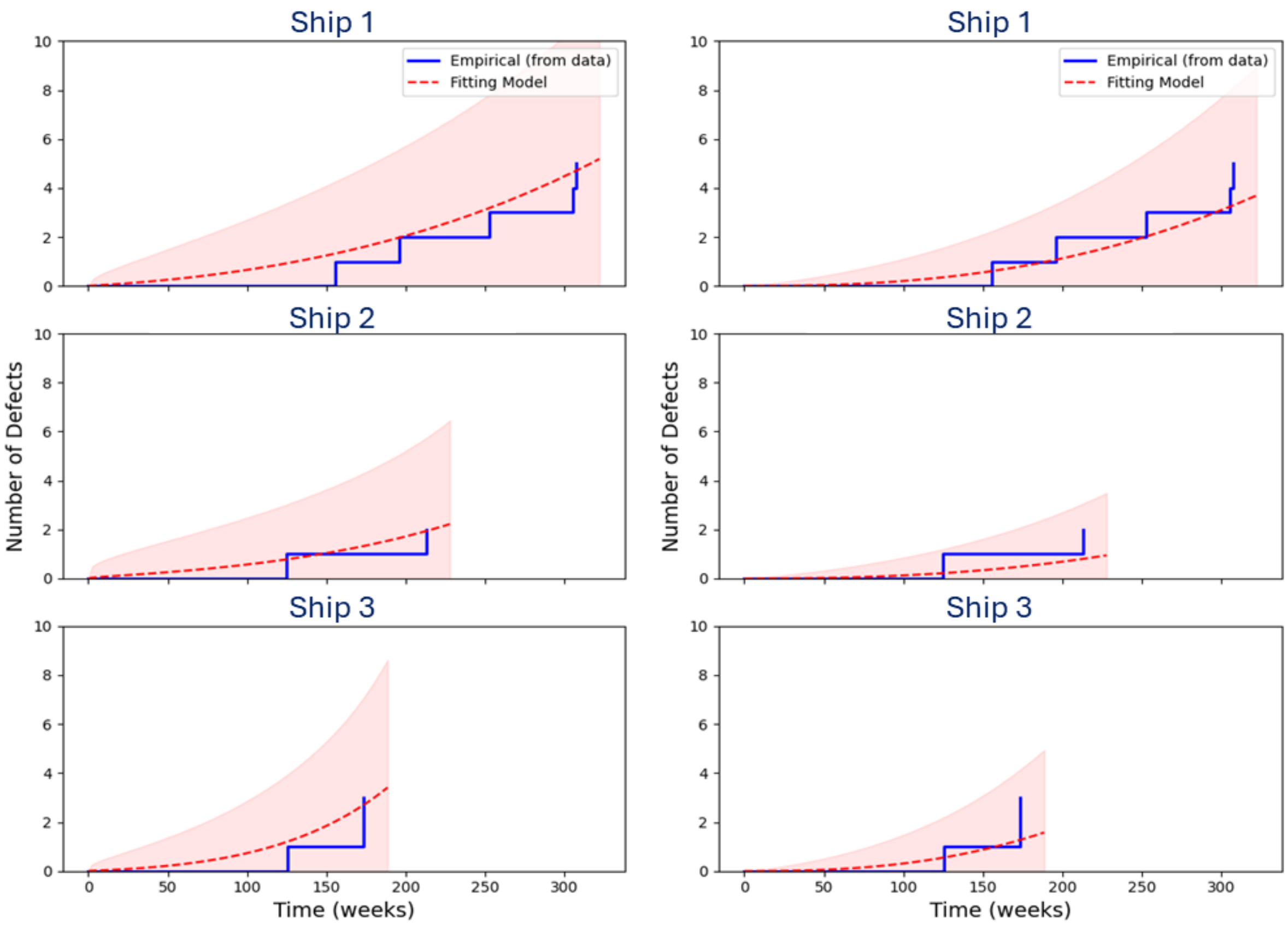

Figure 4: Fitting result vs. actual data: Individual fitting (left) vs. Hierarchical fitting (right)

## 4.3 Validation of Coating Breakdown Prediction

The fitted PL-NHPP model for each compartment allows us to estimate the future number of defect arrivals within a specified interval. These predictions can be extended to new, similar ships by leveraging models built using data from launched vessels. This section details these two prediction approaches and their subsequent validation.

For validation purposes, the defect data is divided into training and testing sets. The training set is used to fit the model, while the testing set assesses its predictive accuracy. Figure 5 provides illustrative examples of predicted new defect arrivals for two compartments, representing scenarios with both sufficient and insufficient data for model fitting. In the case of sufficient data, the prediction exhibits high accuracy, with the testing data point falling within the credible zone. This indicates a high degree of confidence in the prediction. For the case of insufficient data, the prediction remains reasonable, with the testing data point still within the confidence interval. This shows the benefit of hierarchical Bayesian approach that the estimations of other compartments with sufficient data can share to compartments with limited data.

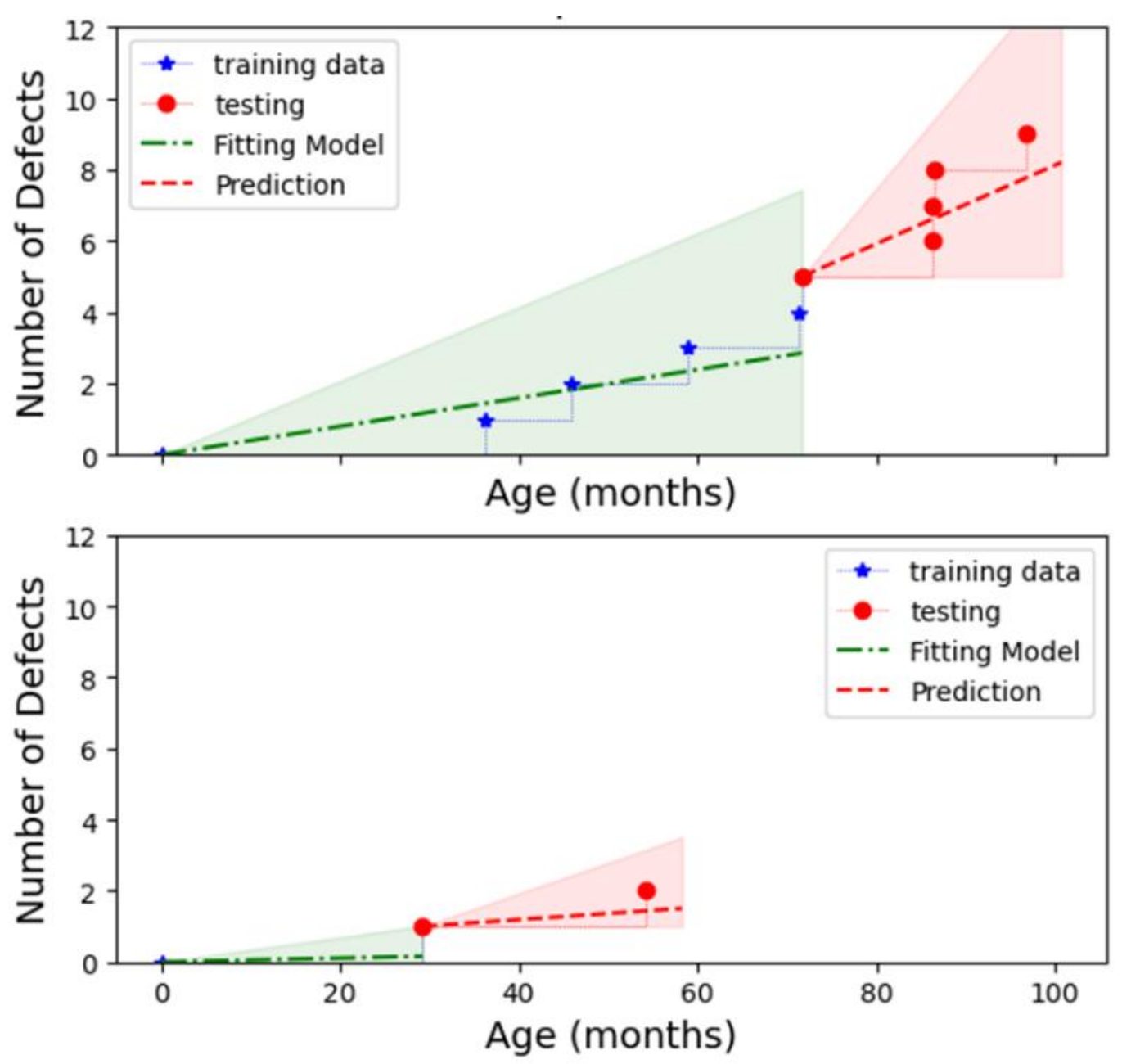

Figure 5: Validation of Arrival Model (Hierarchical fitting): Sufficient data (Above) vs. Insufficient data (Below)

The developed model has the applicability of predicting defect arrivals in a new launched ship (Ship 3). In this scenario, data from Ships 1 and 2 serves as the training set for model fitting, while defect reports from Ship 3 constitute the testing set for validation. Figure 6 presents the predicted defect arrivals for two compartments in Ship 3, representing cases with both sufficient and insufficient data for model fitting in the training set. The results indicate a good fit for PL-NHPP models to data from Ships 1 and 2. The subsequent prediction for Ship 3, achieved by merging these fitted model parameters, demonstrates reasonable accuracy. Even for the case with limited training data, the prediction for the new ship remains acceptable, despite the relative flatness of the PL-NHPP models due to the scarcity of defects observed in the corresponding compartments of Ships 1 and 2. This successful prediction for Ship 3 suggests a high degree of similarity in the coating breakdown patterns between Ship 3 and its sister ships. It is important to note that the model is designed to be continuously updated as new data becomes available. This includes incorporating defect data from the new ship (Ship 3) alongside data from existing ships to further refine the model's predictive capabilities.

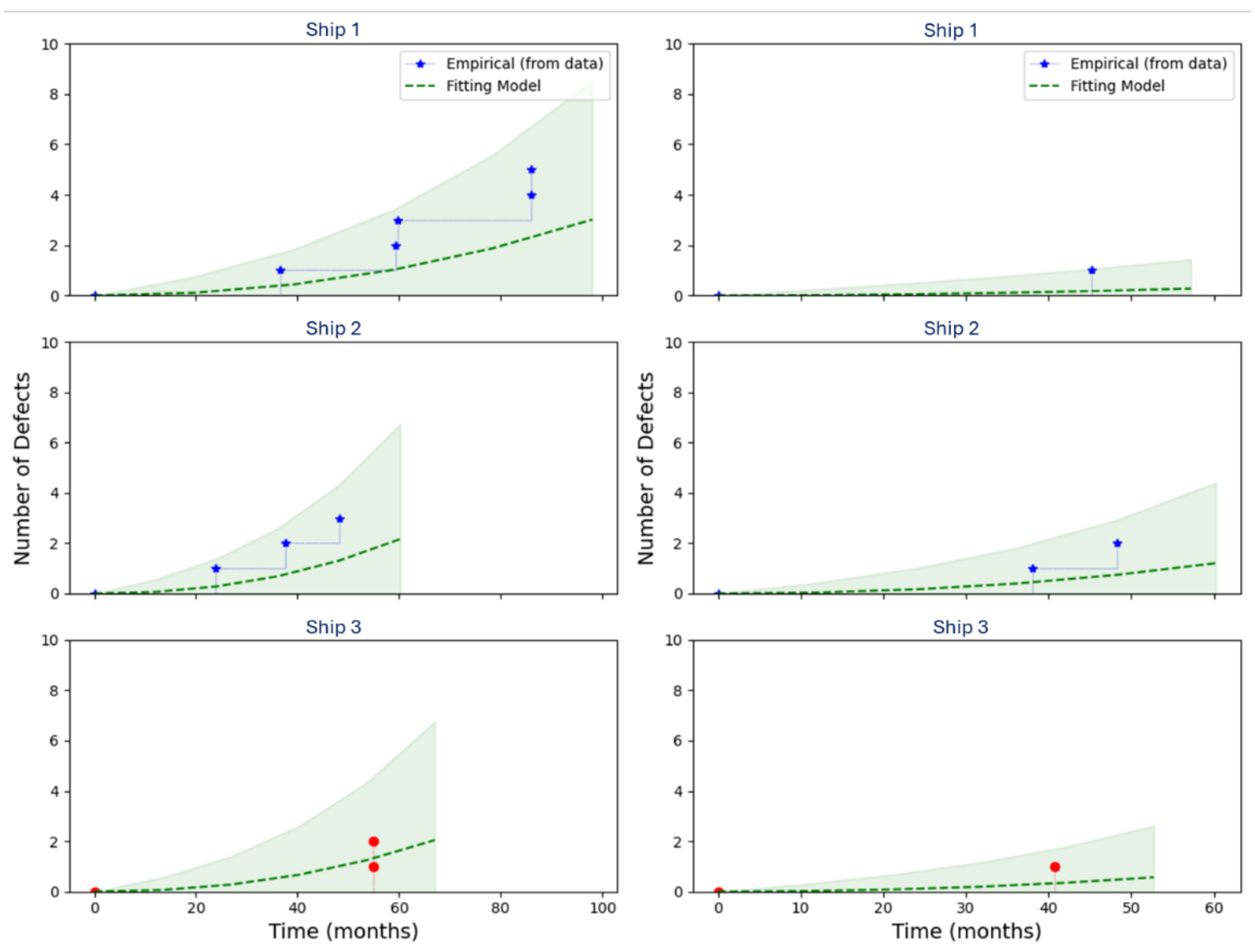

Figure 6: Prediction for new ship: Sufficient data (Left) vs. Insufficient data (Right)

# 5 Inspection Planning for compartments

The defect arrival model facilitates the optimization of inspection plans for compartments, aiming to achieve a balance between the economic burden associated with out-of-service ship time and the repair costs incurred due to delayed maintenance of coating breakdown defects. While cost data limitations in this study necessitate the use of relative costs (Section 3.3), a sensitivity analysis is performed to investigate the adaptability of optimized inspection plans to varying cost structures. The initial cost categories considered are as follows:

| Cost Category | Notation | Value |
|---|---|---|
| **Ship setup cost** | $c_s$ | 500 |
| **Compartment setup cost** | $c_{m,s}^{ins}$ | 10 |
| **Repair cost** | $c_{m,s,\ell}^{rep} = f_c(a_\ell) = \alpha_{m,s} a_\ell^{\beta_{m,s}}$ | $1 \cdot a_\ell^{1.25}$ |

## 5.1 Inspection interval vs. Inspection plan

This study employed a Genetic Algorithm (GA) implemented using the PyGAD package in Python to optimize inspection intervals and plans for all compartments. The population size was set to 3,000, and the algorithm terminated after 10,000 generations without significant improvement in the solution.

Figure 7 presents the distribution of inspection intervals and the corresponding optimized inspection plans for all compartments in Ship 1 over a 20-year period. For comparison, Figure 8 shows the results of the current inspection practices. The results reveal a clustering of inspection intervals, with the majority of compartments scheduled for inspection every 27 months. Interestingly, the inspection plan exhibits a bimodal distribution. Over the 20-year period, a large group of compartments (approximately 75%) undergo inspections around 8 times, while two minor groups experience fewer than 6 inspections. Compared to current inspection practices, the optimized plan results in fewer interval groups and a reduced number of inspections over the 20-year period. This optimization leads to a significantly lower total cost (i.e. about 26% lower: 141k vs. 190k). These findings indicate that the optimized inspection intervals provide considerable benefits compared to current practices.

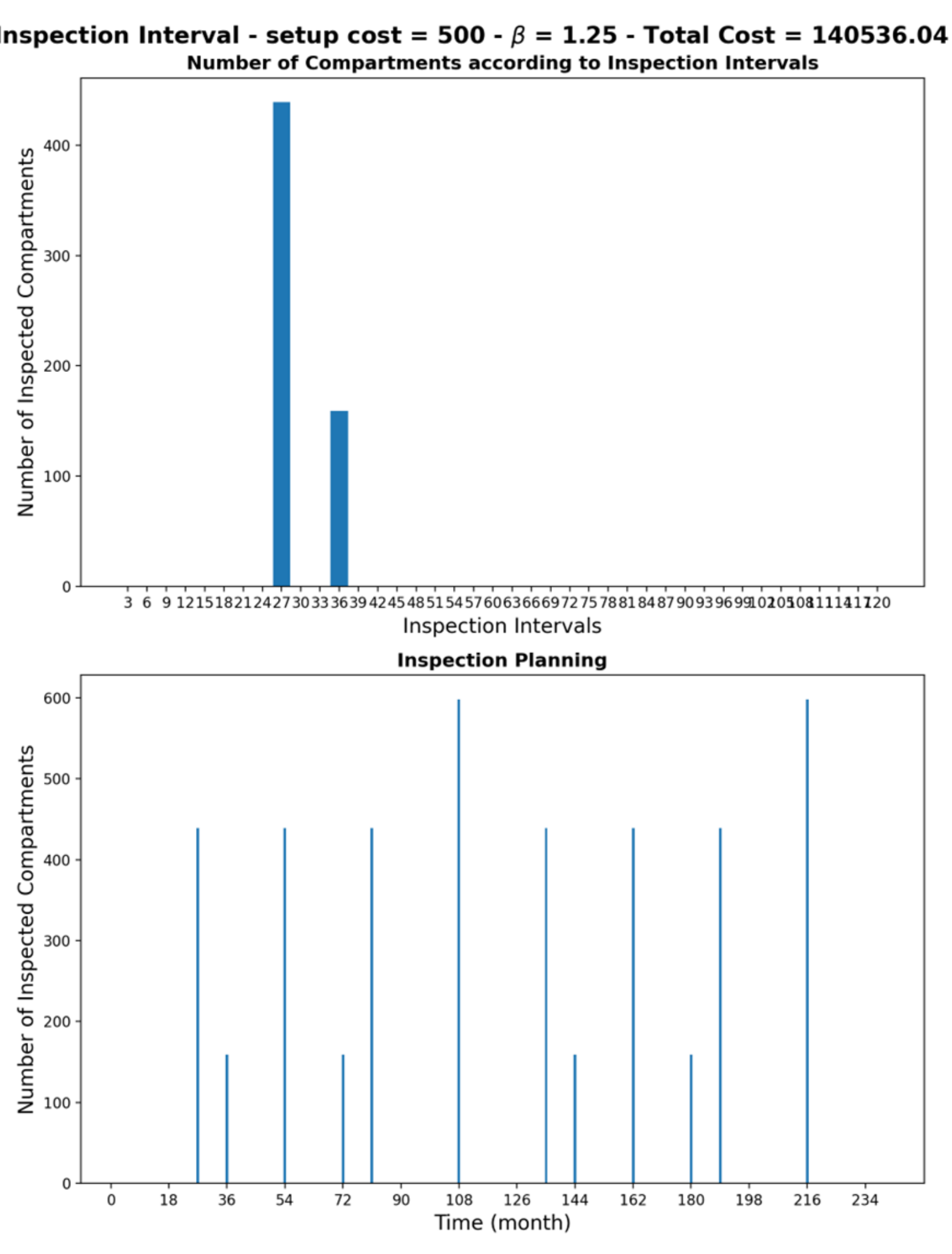

Figure 7: Inspection Interval Optimization: Interval distribution and Inspection planning

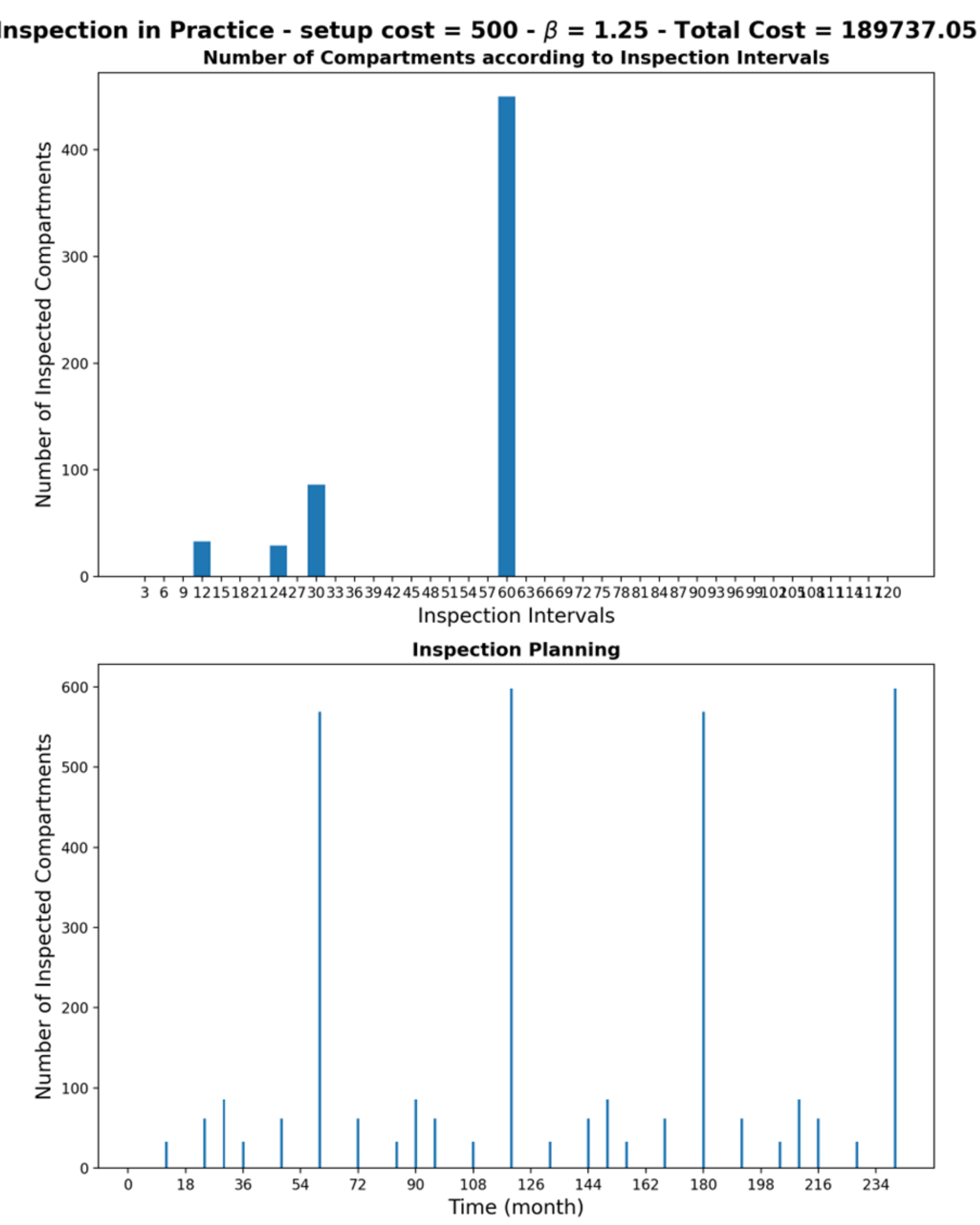

Figure 8: Inspection Planning in Practice: Interval distribution and Inspection planning

Figure 9 depicts the optimized inspection plan derived from the optimization model for a randomly selected subset of compartments. Notably, the plan exhibits a high degree of schedule uniformity, with many compartments sharing identical inspection schedules and consistent intervals between inspections. This observation suggests a strong convergence between the optimized inspection intervals and the optimized inspection schedule. Figure 10 further elucidates this similarity by presenting the total cost and inspection frequency across the scheduling optimization process. While the total cost associated with the optimized plan is lower compared to the outcome of the inspection interval optimization, this is achieved through the flexibility of inspection. Interestingly, the number of inspections and frequency of major inspections (i.e. involving a large number of compartments) remain largely similar between both optimization approaches.

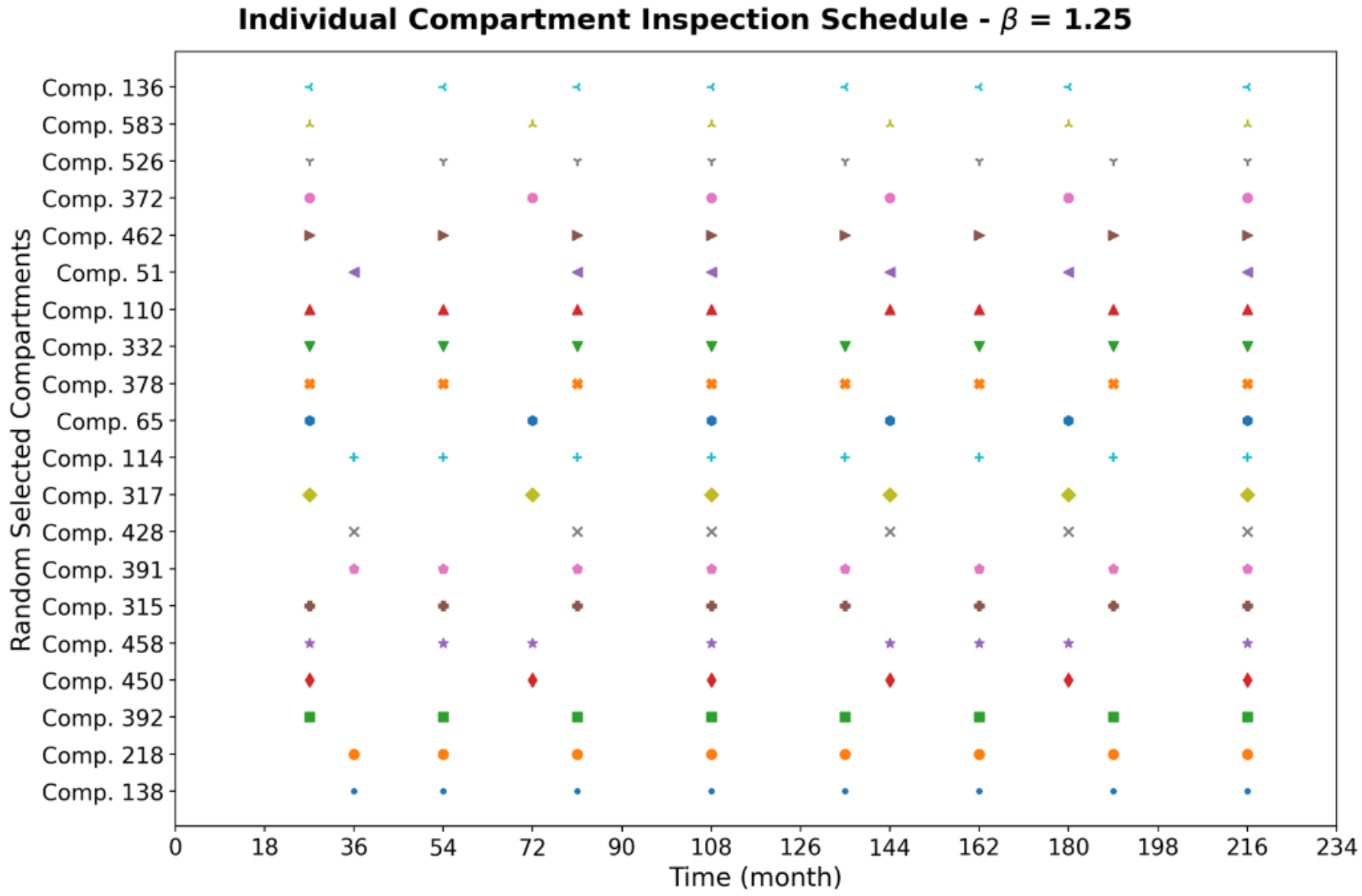

Figure 9: Inspection Schedule Optimization: Inspection schedules of some individual compartments

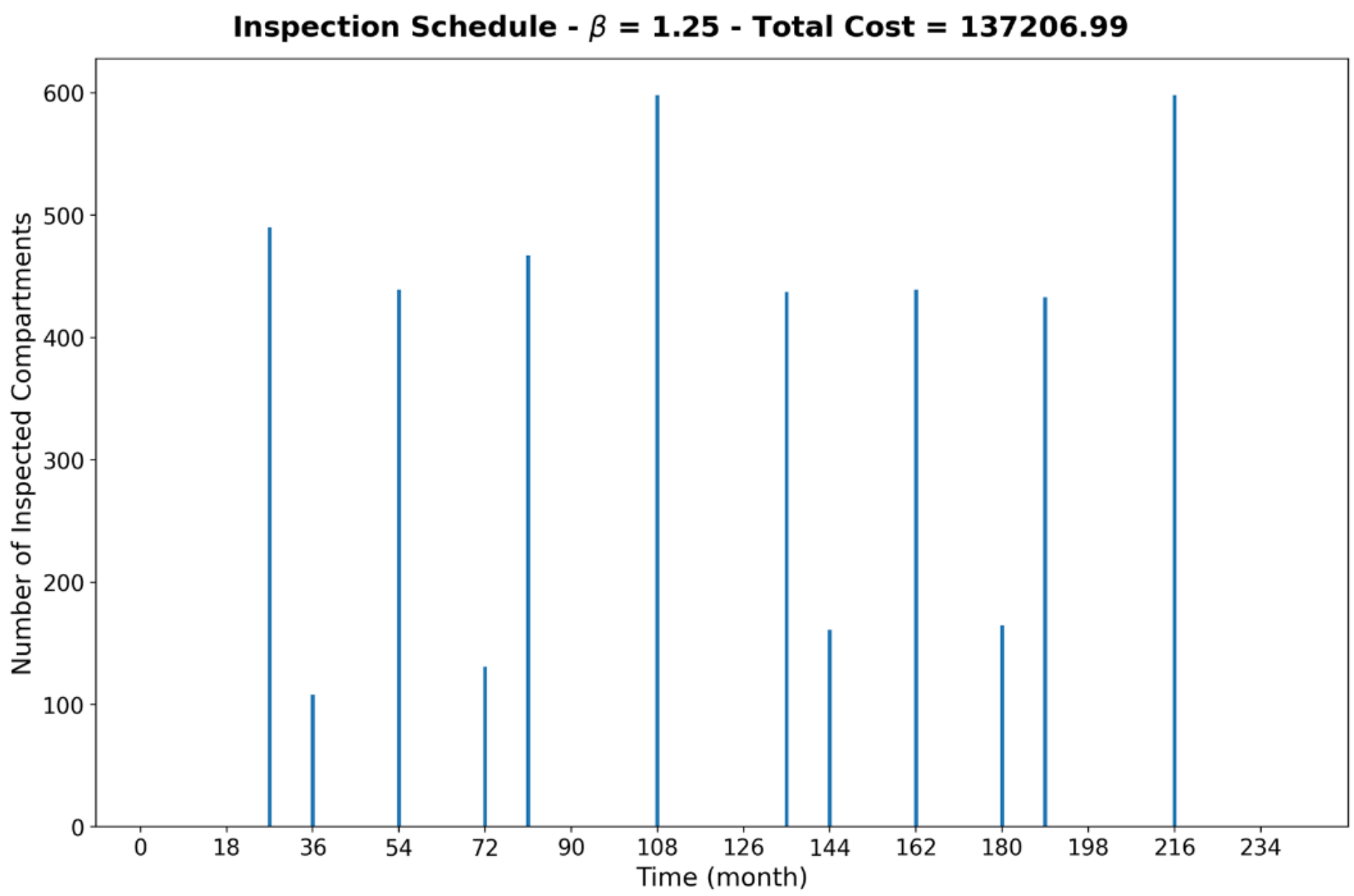

Figure 10: Inspection Schedule Optimization: Total Inspection Schedule

## 5.2 Sensitivity Study

This section evaluates the performance and characteristics of optimized inspection plans under varying cost structures. The comparison is conducted for the current inspection practice and 2 inspection planning optimizations above in which, one is the inspection is planned at a fixed interval time, and the other is flexible schedules for each compartment group. The analysis focuses on the impact of two key parameters: the repair cost pattern, represented by the parameter $\beta_{m,s}$ in Eq. (14), and the out-of-service cost of the ship.

### 5.2.1 Repair cost $c_{m,s,\ell}^{rep}$

Figure 11 illustrates the impact of the repair cost function on the expected total cost of ship repair. The plot demonstrates that a higher value of $\beta$ in the repair cost function results in a steeper increase in repair costs as the defect age increases. Consequently, the total expected repair cost also rises. Notably, the optimized inspection schedule achieves a lower total cost compared to the inspection interval approaches. Both optimized inspection strategies yield substantial savings compared to the inspection in practice. This difference becomes more pronounced as the value of $\beta$ increases, underscoring the cost-saving potential of an optimized schedule in scenarios with rapidly escalating repair costs.

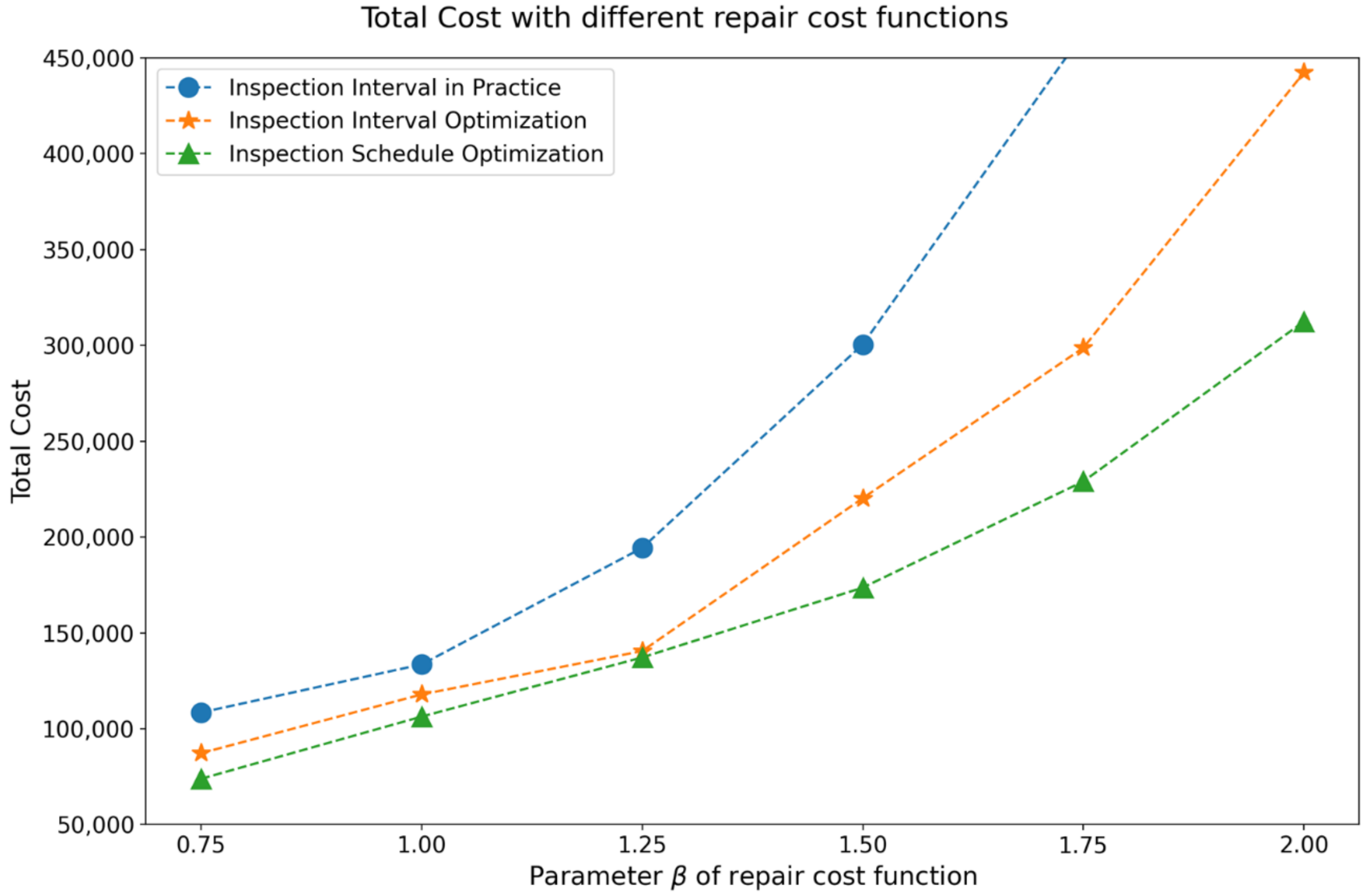

Figure 11: Total Costs with different

The numbers of inspections vs. time of two optimizations, i.e. interval and schedule optimizations, with different $\beta$ are shown in Figure 12 and Figure 13, respectively. In general, there are more inspected compartments and inspection times when $\beta$ increases, and specially, when $\beta < 1$, there are only 5 inspections in schedule optimization. When $\beta$ is larger, the repair cost is more dominant in the cost category, and it increases exponentially to the delay time of repair. Thus, logically, the more frequent inspections are, the shorter delay time for repairs will be, thus, keeping repair costs low. Comparing two optimization approaches, the schedule optimization results in fewer inspections and a reduced number of inspections at the early age of the ship.

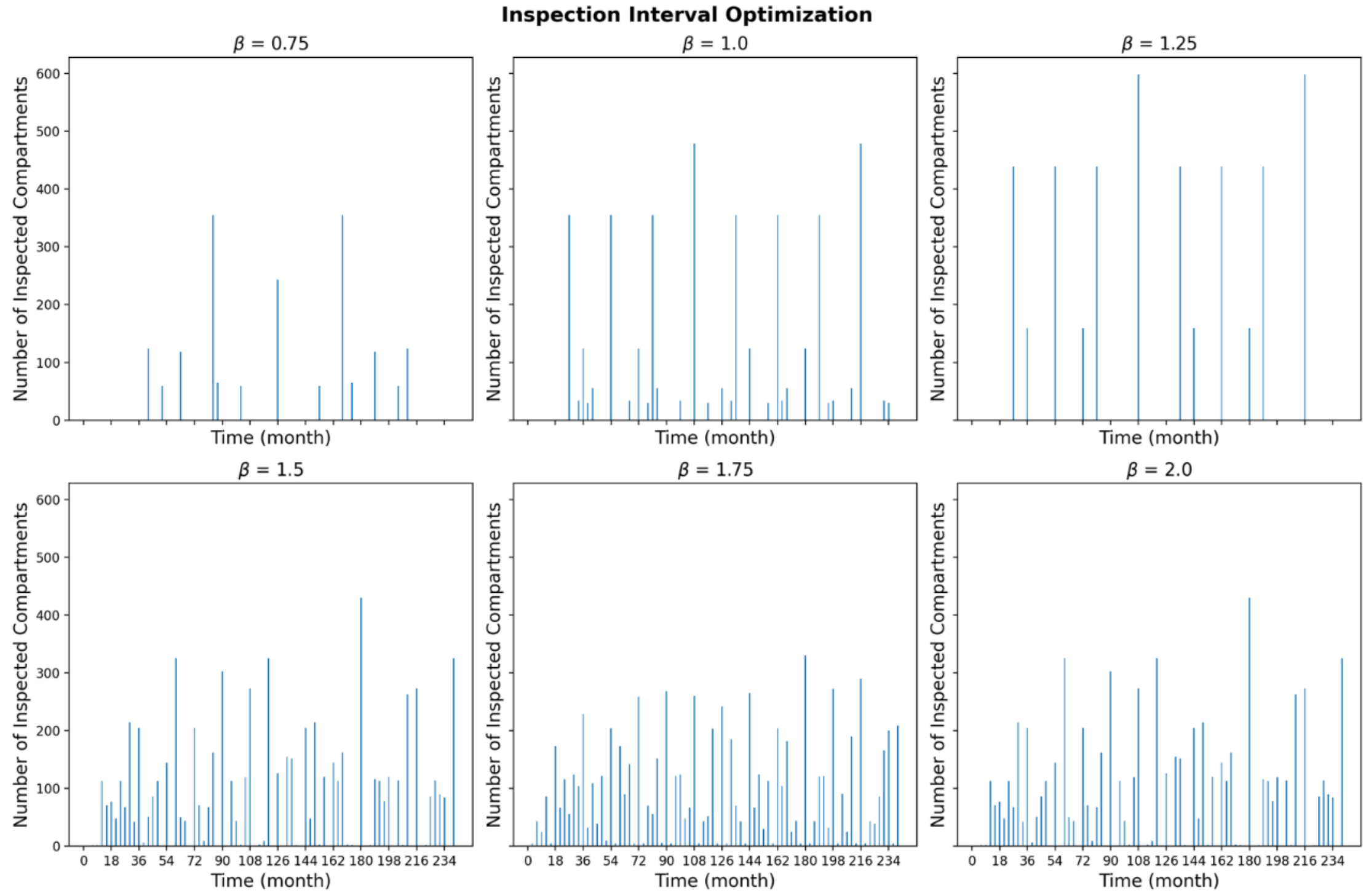

Figure 12: Numbers of Inspections according to $\beta$: Inspection Interval Optimization

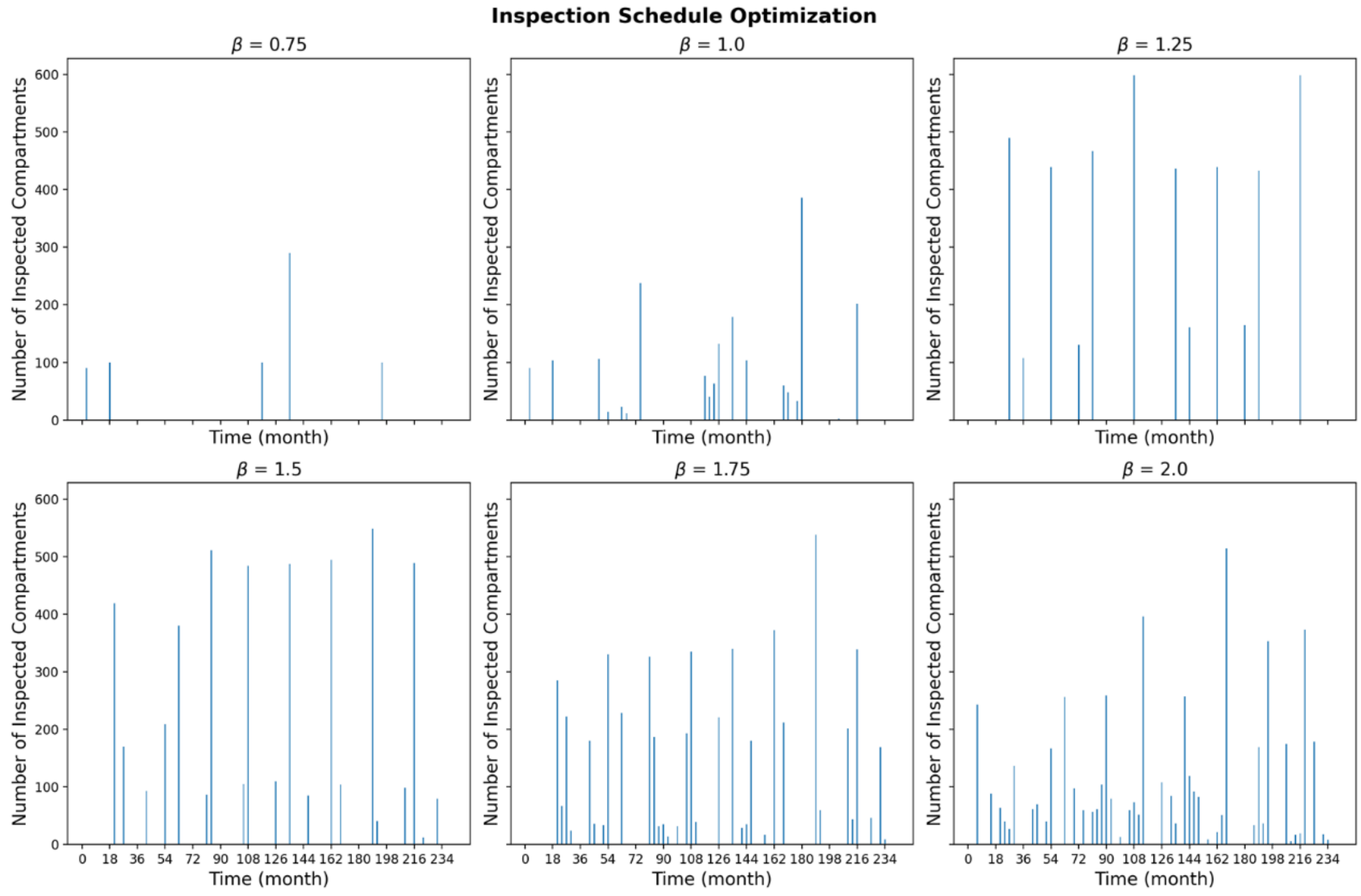

Figure 13: Numbers of Inspections according to $\beta$: Inspection Schedule Optimization

### 5.2.2 Out-of-service cost $c_s$

Figure 14 depicts the influence of the out-of-service cost on the expected total cost of ship repair. The plot reveals that a higher value of $c_s$ in the repair cost function leads to an increase in repair costs. Notably, the optimized inspection schedule achieves a lower total cost compared to the inspection interval approaches. This gap between fixed interval inspection in practice and two optimized inspection planning become more significant as the value of $c_s$ increases, except when this cost is too big, highlighting the cost-saving potential of the optimized schedule under scenarios with rapidly escalating repair costs.

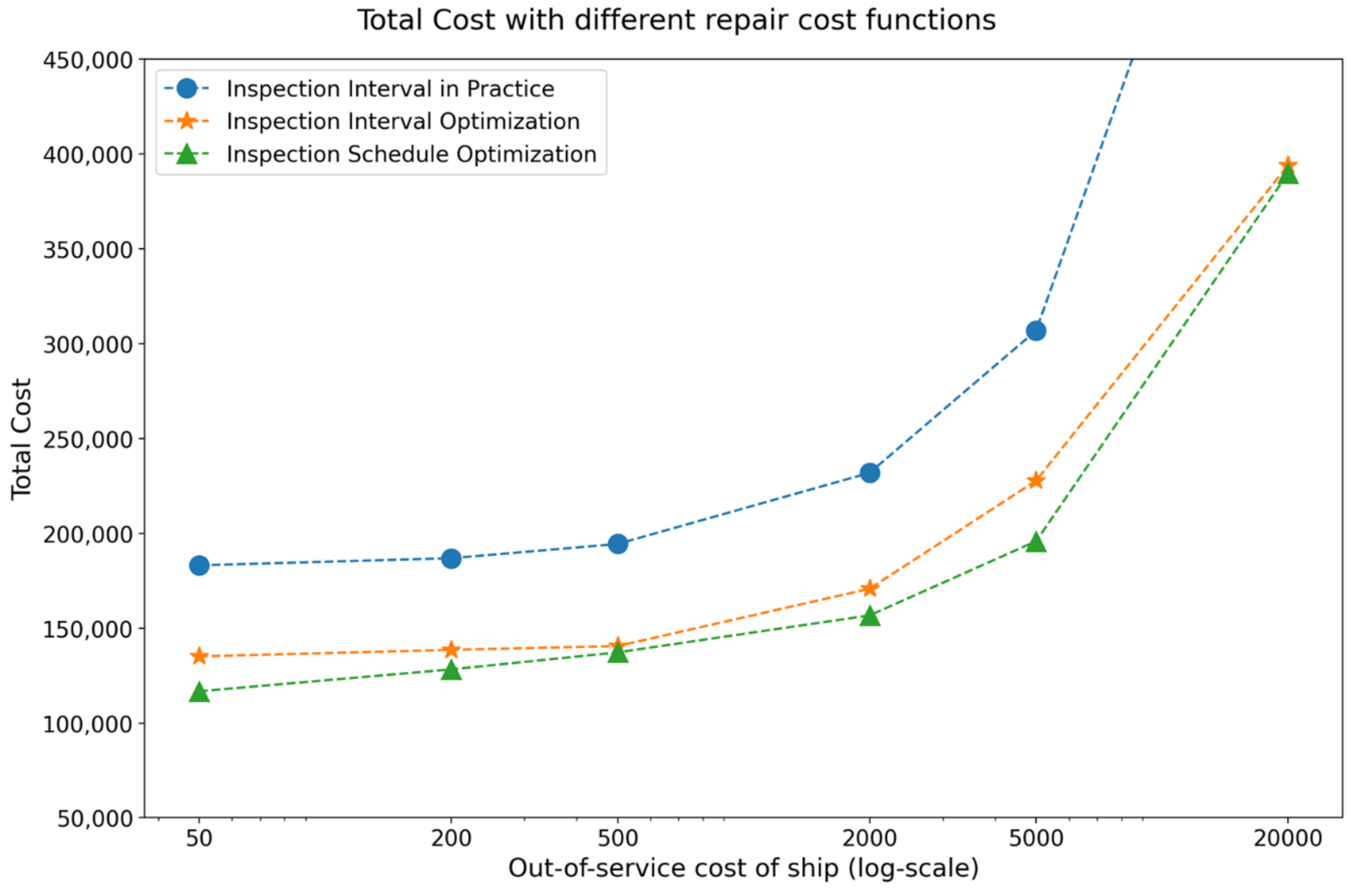

Figure 14: Total Costs with different $c_s$

The numbers of inspections vs. time of two optimizations, i.e. interval and schedule optimizations, with different $c_s$ are shown in Figure 15 and Figure 16, respectively. In general, there are more inspection times when $c_s$ is small. When $c_s$ is larger, the out-of-service cost is more dominant in cost category, and ships therefore is delayed for inspection. On the other hand, if the out-of-mission cost of ships is small, the repair cost or existing time of defects is more important in inspection decision. Comparing two optimization approaches, the schedule optimization results in little fewer numbers of inspections by grouping more compartments in an inspection time. However, this difference is small which is relevant to the small cost gaps between two optimization approaches shown in Figure 14 above.

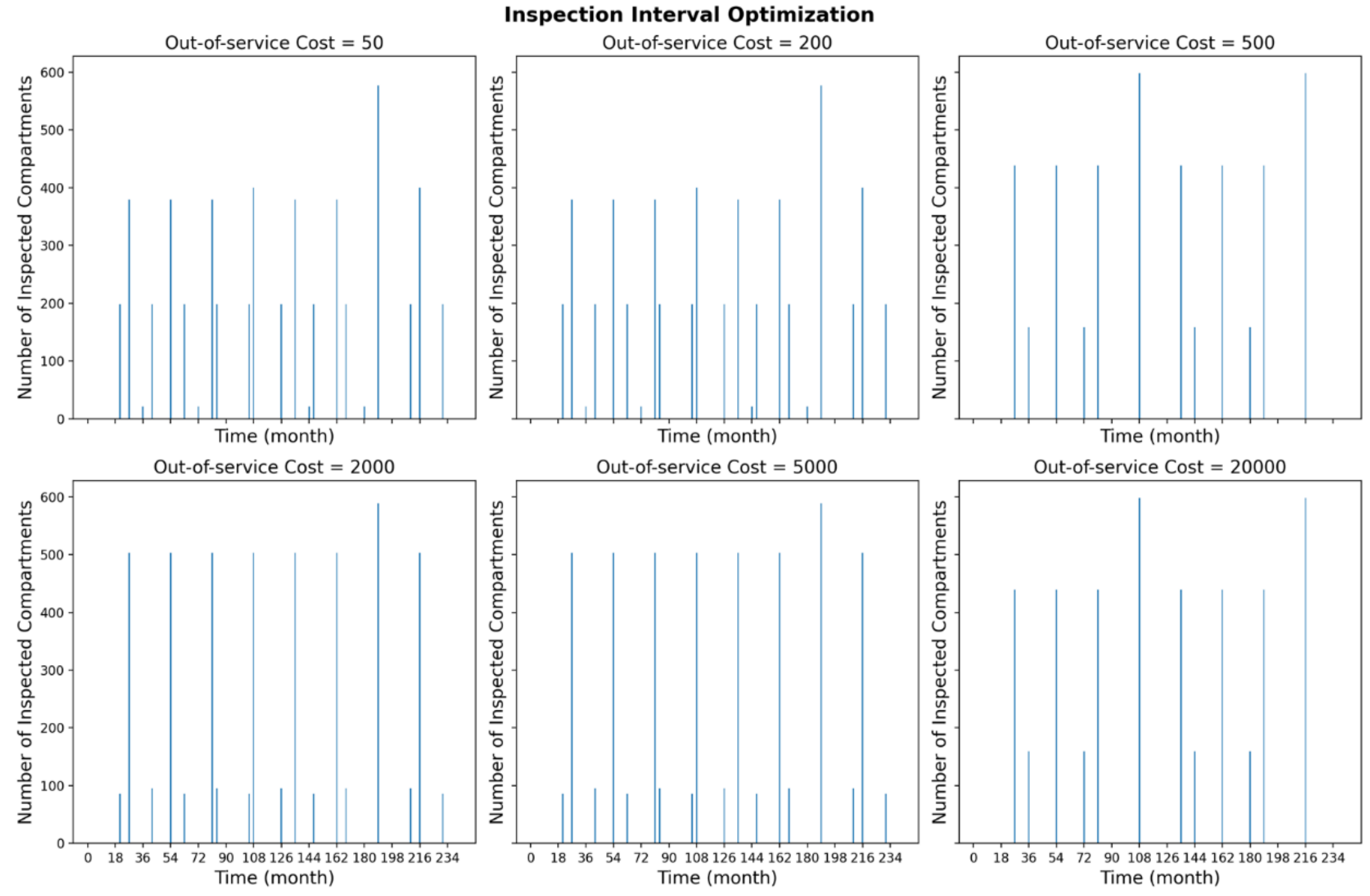

Figure 15: Numbers of Inspections according to $c_s$: Inspection Interval Optimization

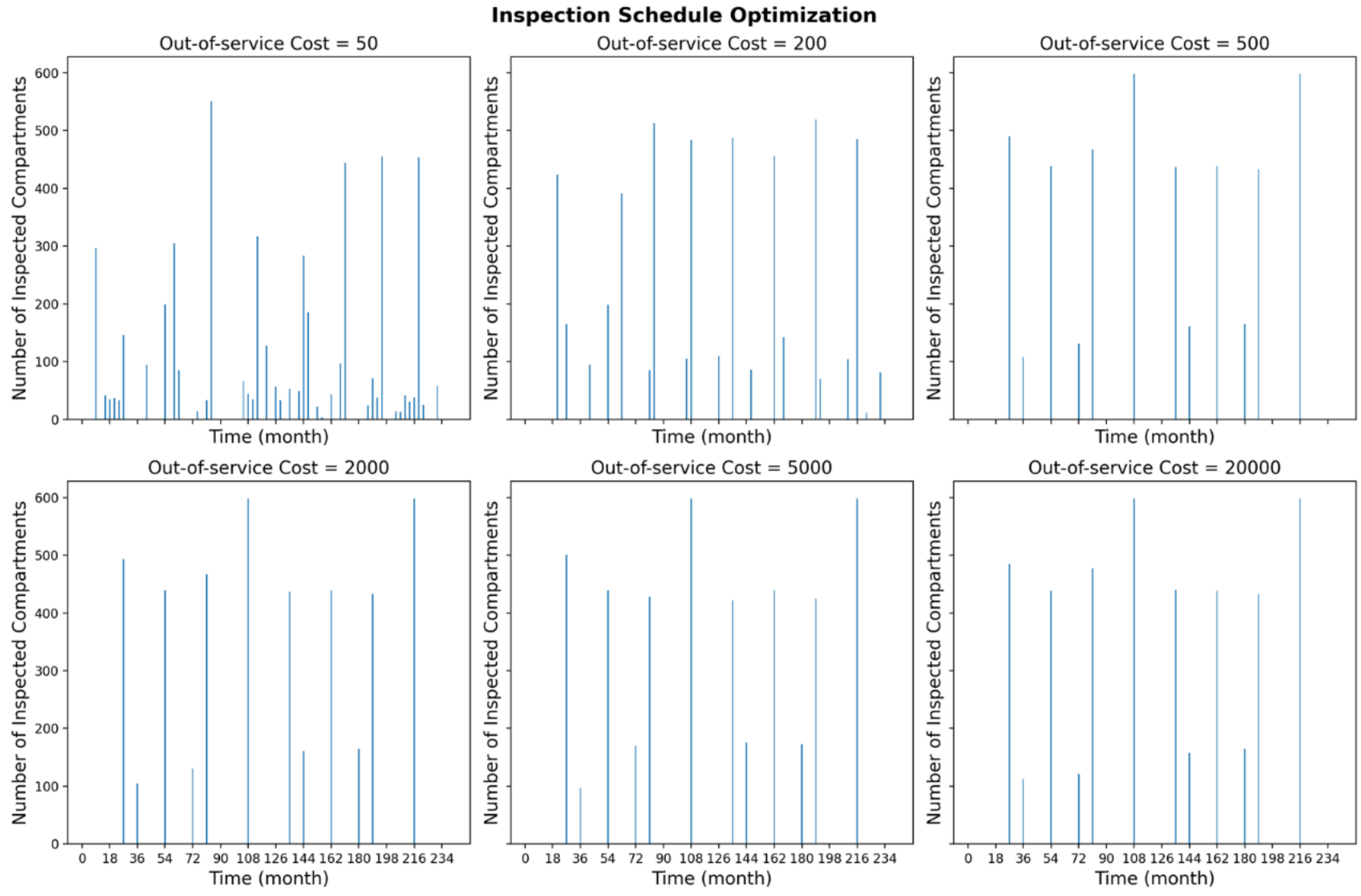

Figure 16: Numbers of Inspections according to $c_s$: Inspection Schedule Optimization

# 6 Conclusion

This study proposes a methodology utilizing a hierarchical Bayesian approach to estimate the parameters of an arrival model for coating breakdown defects on vessels. The hierarchical structure effectively addresses limitations associated with scarce/insufficient data, enabling robust model fitting and the generation of credible intervals.

Furthermore, the research develops optimization models for ship inspection planning through two distinct approaches: inspection interval optimization and inspection scheduling optimization. These models incorporate repair costs (influenced by the random exposure time of undetected defects) and out-of-service costs due to inspections. Sensitivity analyses are employed to comprehensively evaluate these factors.

The methodology and models were validated utilizing a real-world case study with data from three marine vessels. The models exhibited exceptional performance in forecasting defect arrivals across the entire fleet, thus providing valuable risk assessments for maintenance decision-makers. Furthermore, the optimization of inspection planning successfully identified plans that achieve an optimal balance between coating and corrosion defect repair costs and vessel downtime. Comparisons between two inspection planning approaches—inspection scheduling versus fixed inspection interval planning for each individual compartment—revealed that although fixed interval planning is not as effective as scheduling, it remains a viable option due to its simplified implementation and minimal performance loss, especially in scenarios where the cost factor related to defect severity due to repair delay is low.

Future research endeavors will concentrate on modeling the severity progression of coating breakdown defects due to delayed repairs. This model will serve as the foundation for developing and optimizing condition-based or prediction-based maintenance policies for ships.

## Acknowledgement

The authors wish to express our sincere thanks for the invaluable support given by Alison Court and Mr. Kumar Shobhit.

# Appendix

$$\bar{A}_k(t_1,t_2) = \int_{t_1}^{t_2} (t_2 - t) f_{k,t_1}(t)dt = t_2 F_{k,t_1}(t_2) - \int_{t_1}^{t_2} t f_{k,t_1}(t)dt$$

Assign $S = \int_{t_1}^{t_2} t f_{k,t_1}(t)dt = \int_{t_1}^{t_2} \frac{\left[a\left(t^b - t_1^b\right)\right]^{k-1}}{(k-1)!}(abt^b)e^{-a\left(t^b - t_1^b\right)}dt$

Let $y = at^b \rightarrow dy = abt^{b-1}$ and $t = \left(\frac{y}{a}\right)^{\frac{1}{b}}$

$$S = \frac{e^{at_1^b}}{(k-1)!} \int_{at_1^b}^{at_2^b} \left(y - at_1^b\right)^{k-1} \left(\frac{y}{a}\right)^{\frac{1}{b}} e^{-y} dy$$

$$S = \frac{e^{at_1^b}}{(k-1)!} \int_{at_1^b}^{at_2^b} \left[\sum_{i=0}^{k-1} \binom{k-1}{i} \left(-at_1^b\right)^i y^{k-1-i}\right] \left(\frac{y}{a}\right)^{\frac{1}{b}} e^{-y} dy$$

$$S = \frac{e^{at_1^b}}{a^{\frac{1}{b}}(k-1)!} \sum_{i=0}^{k-1} \binom{k-1}{i} \left(-at_1^b\right)^i \int_{at_1^b}^{at_2^b} y^{k-1-i+\frac{1}{b}} \cdot e^{-y} dy$$

$$S = \frac{e^{at_1^b}}{a^{\frac{1}{b}}(k-1)!} \sum_{i=0}^{k-1} \binom{k-1}{i} \left(-at_1^b\right)^i \left[\gamma\left(k - i + \frac{1}{b}, at_2^b\right) - \gamma\left(k - i + \frac{1}{b}, at_1^b\right)\right]$$

With $F_{k,t_1}(t_2) = \frac{\gamma\left(k, a\left(t_2^b - t_1^b\right)\right)}{(k-1)!}$, the final result will be:

$$\bar{A}_k(t_1, t_2) = \frac{t_2 \gamma\left(k, a\left(t_2^b - t_1^b\right)\right)}{(k-1)!} - \frac{e^{a t_1^b}}{a^{\frac{1}{b}}(k-1)!} \sum_{i=0}^{k-1} \binom{k-1}{i} \left(-a t_1^b\right)^i \left[\gamma\left(k - i + \frac{1}{b}, a t_2^b\right) - \gamma\left(k - i + \frac{1}{b}, a t_1^b\right)\right]$$